\documentclass[]{interact}

\usepackage{epstopdf}
\usepackage{subfigure}
\usepackage{graphicx}
\usepackage{xcolor}
\usepackage{soul}

\usepackage{amsmath}
\usepackage{amssymb}
 \usepackage{url}

\usepackage[square,numbers]{natbib}
\theoremstyle{plain}

\theoremstyle{definition}

\theoremstyle{remark}

\begin{document}


\title{The physics of learning machines.}

\author{
\name{G.~J. Milburn\textsuperscript{a}\thanks{ milburn@physics.uq.edu.au} and Sahar Basiri-Esfahani\textsuperscript{b}} 
\affil{\textsuperscript{a}Centre for Engineered Quantum Systems, School of Mathematics and Physics, The University of Queensland, St Lucia QLD 4072, Australia;}
    \textsuperscript{b}Department of Physics, Swansea University, Singleton Park, Swansea SA2 8PP,
Wales, United Kingdom}

\maketitle

\begin{abstract}
A learning machine, like all machines,  is an open system driven far from thermal equilibrium by access to a low entropy source of free energy. We discuss the connection between machines that learn, with low probability of error, and the optimal use of thermodynamic resources for both classical and quantum machines.  Both fixed point and spiking perceptrons are discussed in the context of possible physical implementations. An example of a single photon quantum kernel evaluation illustrates the important role for quantum coherence in data representation.  Machine learning algorithms, implemented on conventional complementary metal oxide semiconductor (CMOS) devices, currently consume large amounts of energy. By focusing on the physical constraints of learning machines rather than algorithms, we suggest that a more efficient means of implementing learning may be possible based on quantum switches operating at very low power. Single photon kernel evaluation is an example of the energy efficiency that might be possible.
\end{abstract}


\section{Introduction}
 As everyone knows, the development of machine learning algorithms, such as deep neural networks, has had a huge impact on technology~\cite{Goodfellow}. This paper is concerned with something quite different; machines that learn.  Learning machines are analogue physical devices that learn. Deep neural networks are algorithms. In learning machines the algorithms are written by the laws of physics, especially thermodynamics.  Viewing machine learning from a physical perspective is not new; algorithmic neural networks were inspired by physical learning in the mammalian brain.  Our aim in this paper is to discuss what kinds of machines can learn and how they are constrained by classical and quantum physics. We do not discuss how to use machine learning algorithms to  solve physics problems~\cite{Karniadakis,RevModPhys.91.045002} or configure experiments~\cite{Craig2021}, nor do we attempt to give physical analogies for deep learning algorithms.    

 All machines are constrained by the laws of physics, especially thermodynamics. Of course, given a sufficiently accurate description of their construction, their dynamics may be simulated on a digital computer. But a simulated learning machine is no more a machine than a simulated hurricane is a hurricane. In both cases, the machine behind the algorithm --- typically vast networks of coupled transistors -- is the machine.   Deep neural networks are simulations of something related, very distantly, to a model for neurons in the mammalian brain; a rather special kind of machine. 
 
A machine is an open system driven far from thermal equilibrium by access to a low entropy source of free energy. In biological machines, that source is the sun. In mobile phones, it is a battery. 
As in any open system, a machine dissipates  heat and produces entropy. It is an irreversible device. This much is true of learning machines, but the fluctuations that accompany dissipation turn out to be critical for machines to learn.

A central issue for the physics of learning machines is the trade-off between the learning rate
and power dissipated. This is also a central issue for deep learning algorithms running on silicon processors. Chojnnacka~\cite{Chojnnacka} summarises a number of studies of the current costs in terms of their carbon footprint.  As an example, a natural language processing application using a transformer deep learning model with neural architecture search 'costs' over 600,000 lbs of $CO_2$ to train. This is 17 times the $CO_2$ emissions of an average American over one year. It seems unlikely that deep learning is making optimal use of scarce thermodynamics resources. Is there a minimum thermodynamic cost to learning machines?   

\begin{figure}[t!]
    \centering
    \includegraphics[scale=0.4]{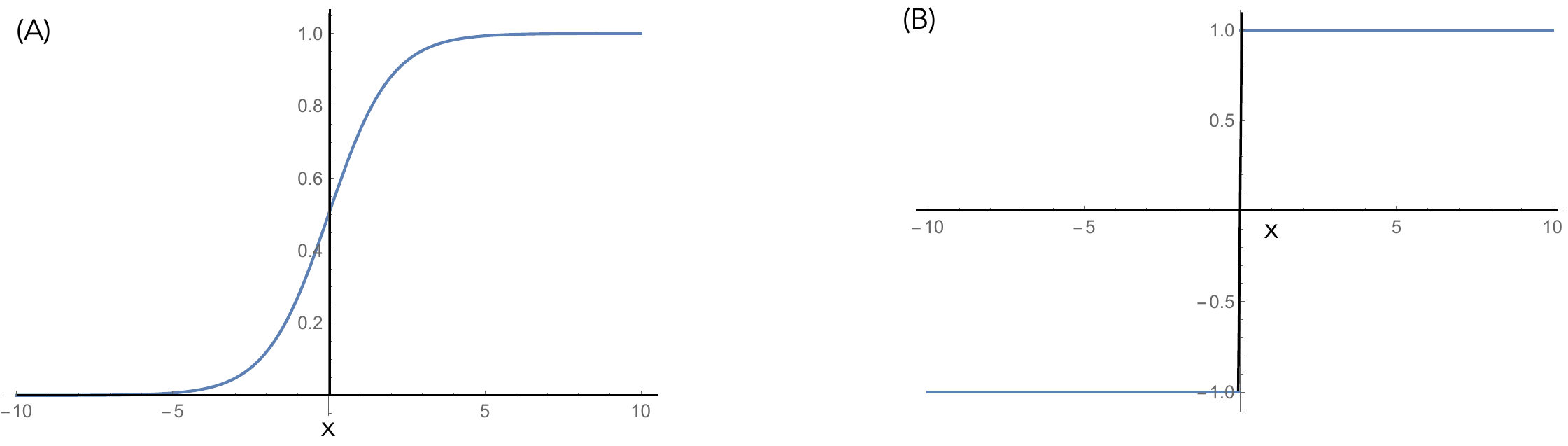}
    \caption{Two examples of an activation function. (A) The sigmoidal function, and (B) the sign function.}
    \label{af}
\end{figure}
In neural network algorithms, a key role is played by the {\em activation function}, a real valued, non-linear function of its inputs, $\vec{\xi}=(\xi_0, \xi_1,\xi_2,\ldots \xi_n)$. 
Activation function assigns an output to the weighted sum of an input to the neuron to decide what should be input to subsequent neurons in the network.
A typical example is the sigmoidal activation function
\begin{equation}
    f_w(\vec{\xi}) = (1+e^{-\beta \vec{\xi}\cdot\vec{w}})^{-1},
\end{equation}
where $\beta$ is a positive real number, $\vec{w}=(w_0, w_1, w_2, \ldots w_n)$ are called the {\em weights}, and we assume $\xi_0$ is fixed so that $w_0\xi_0$ is the {\em bias}. Other choices may be made. For example, if we require the output to be a binary number we can choose the sign function
\begin{equation}
    f_w(\vec{\xi}) = (1+\mbox{sign}(\vec{\xi}\cdot\vec{w}))/2,
\end{equation}
where 
\begin{equation}
    \mbox{sign}(z) = \left \{\begin{array}{ll}
                            +1 &  if \ \ \  z > 0,\\
                            -1 & if \ \ \   z\leq 0.\end{array}\right.
\end{equation} 
In Fig.~(\ref{af}), we plot these two functions. 
These functions are mathematical algorithms that describe the action of a switch. As we will see, the sigmoidal function describes the average value of the  output in a stochastic digital physical switch.

\section{Learning machines versus machine learning} 
\label{LM-ML}
The transition to learning machines is made by replacing an algorithm to evaluate an activation function,  with an actual physical device that implements the switch. We will call this an {\em activation switch} to distinguish it from the activation function.  In the physical switch, the weights become physical bias/driving forces that change the physical state of the switch. However, there is now a crucial distinction we need to make between the average behaviour  of the switch as a function of the bias, averaged over many trials, and the {\em stochastic} response of the switch from one trial to the next. This distinction necessarily enters as an activation switch is a dissipative device and hence, by the fluctuation-dissipation theorem, must be accompanied by noise. We then find that the activation function gives the \emph{probability} of the output --- a binary number --- instead of giving the output or the predicted value. It describes the stochastic response of the activation switch to a change in the  bias forces. 

\begin{figure}
     \centering
     \includegraphics[scale=0.5]{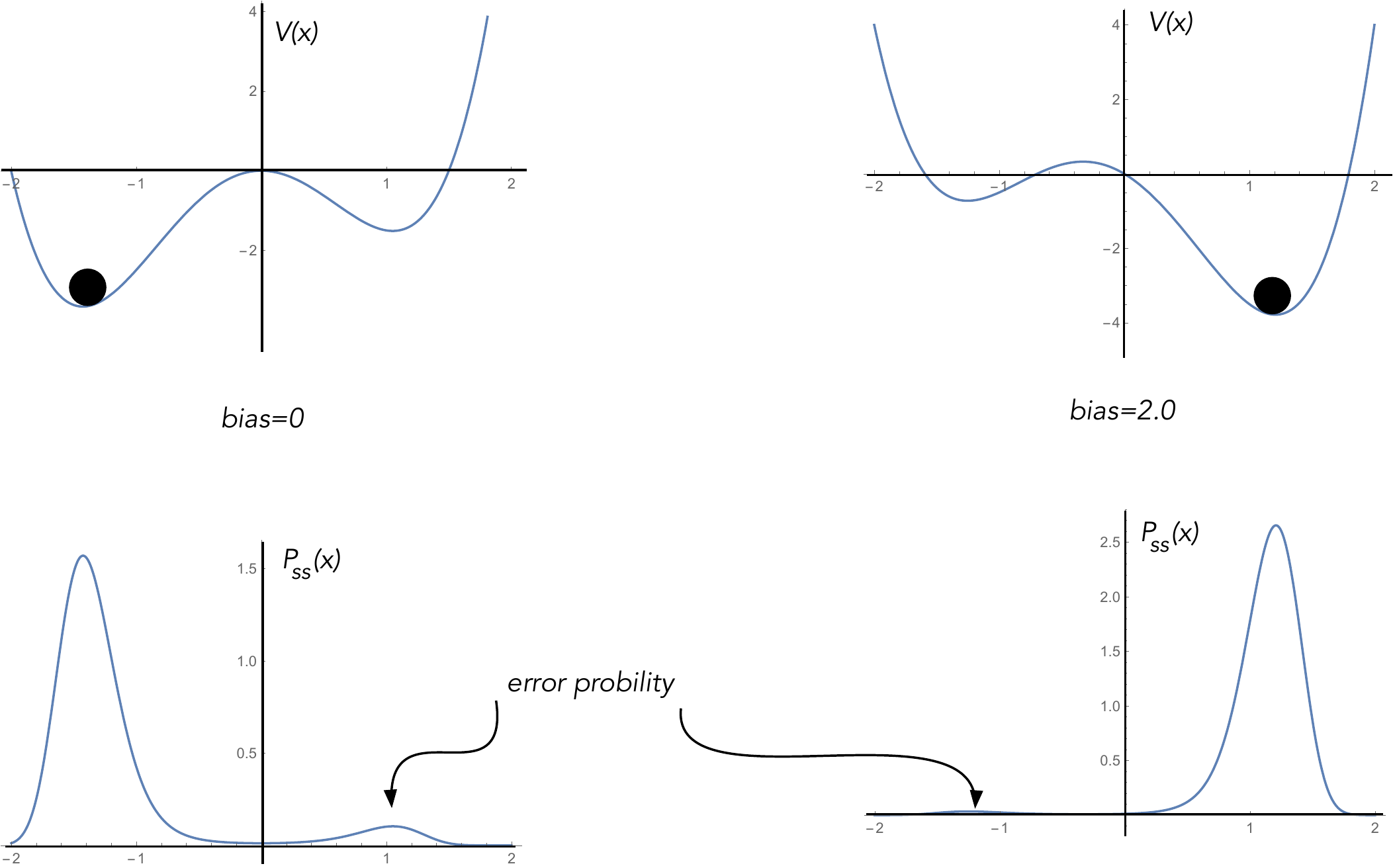}
     \caption{An example of an activation switch based on a particle moving with high friction in a double-well potential.  The output variable is simply the sign of the displacement, $x(t)$, of the particle.  (Top) Application of a bias force changes the shape of the potential so that the lowest energy state changes from negative to positive displacement.  This requires work to be done on the particle, work that is dissipated as heat. (Bottom) The steady state probability distribution when noise is included obtained by solving the Smoluchowski equation. Note that there is a small probability of finding the system in the wrong position.   }
     \label{double-well}
 \end{figure}

We can illustrate this distinction with a simple dynamical system based on a particle moving in a double well potential with high friction and coupled to a thermal bath at temperature $T$, see Fig.~(\ref{double-well}).
The stochastic dynamics of this system is described by the Smoluchowski equation (see \citep{Gardiner} Chapter 6). Depending on the sign of a linear bias potential, the system has a steady state localised on the left or the right of the potential barrier.  The steady state however is probabilistic as thermal fluctuations can drive the system from one well to the other.  The steady state probability distribution reflects this by having finite support in both wells.  This results in a small error probability that the system will be found in the wrong state for a given bias.  

\begin{figure}
    \centering
    \includegraphics[scale=0.5]{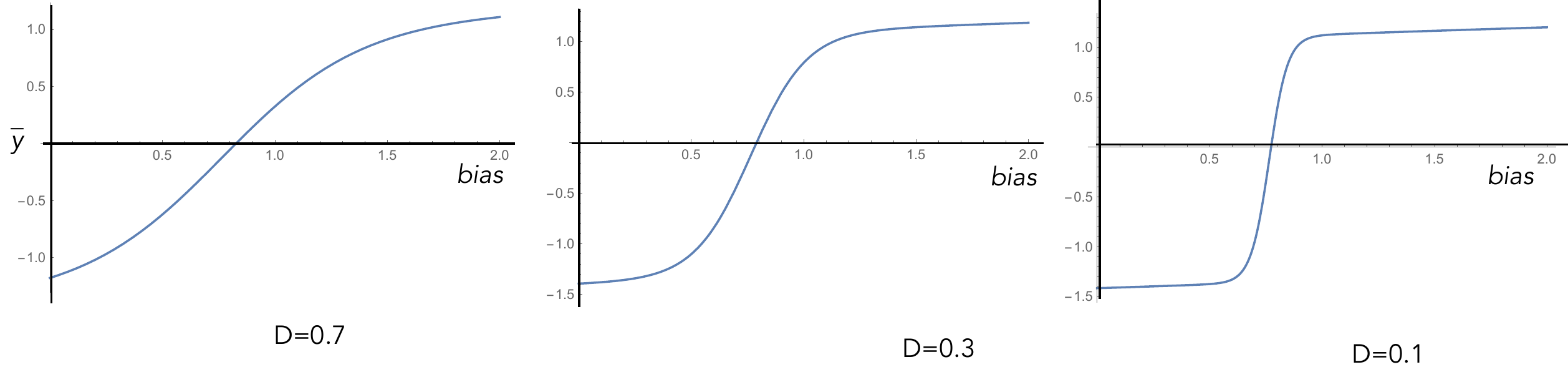}
    \caption{The steady-state ensemble average of the mean displacement of the particle in the double well versus the bias with varying noise levels (i.e.temperature).The noise is parameterised by a diffusion constant $D$ that is proportional to the temperature by the fluctuation-dissipation theorem.  On the left the noise is high and the switching nature of the relationship between output and input is unclear. As the noise is decreased a more definite switch is seen as a function of the bias of the potential.  }
    \label{DW-mean}
\end{figure}

\begin{figure}
    \centering
    \includegraphics[scale=0.5]{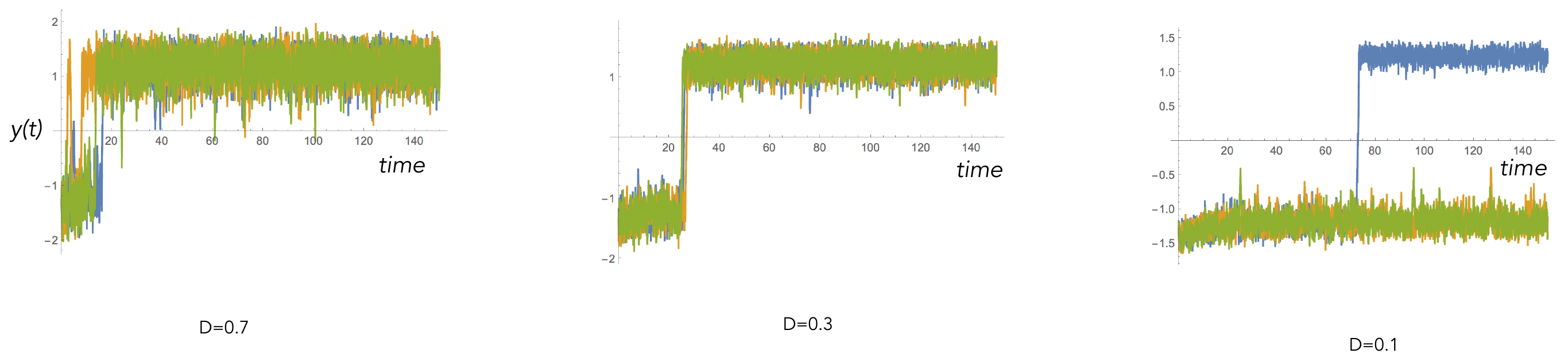}
    \caption{Sample trajectories of the displacement of the particle in the double well activation switch as a function of time as the noise is decreased from left to right. In each case there are three samples $y(t)$.  The bias of the potential was increased from $0$ to a constant maximum value. When the noise is large (left) the output switches quickly but the stochastic fluctuations are large. When the noise is small (right) in only one sample did the device switch at all implying an error on the two that did not switch.  Further more as the noise is decreased the switch is slower. In the limit of no noise there is no switching at all.  Dissipation and noise are essential for efficient learning machines.  These plots were generated by solving the Ito stochastic differential equation for the Smoluchowski process in \emph{Mathematica}.   }
    \label{DW-stochastic}
\end{figure}

One way to view the behaviour of the device is in terms of the steady state mean value of the physical quantity that distinguishes one state from another. This can be calculated as  a function of the linear bias gradient, see Fig.~(\ref{DW-mean}). 
The steady state description in terms of a stationary probability distribution is one way to describe this system. Equally important is a description in terms of its stochastic dynamics. In Fig.~(\ref{DW-stochastic}), we consider the dynamics of the physical quantity that distinguishes one state from another.
In a single trial, the output is in fact a stochastic variable,  with the size of the dissipation determining the scale of the fluctuations, as measured by the diffusion constant $D$. As the diffusion constant is proportional to the rate of energy dissipation times the temperature, the lower the temperature the lower the noise.  This means that in some cases the output will not switch when it should, corresponding to an error, see Fig.~(\ref{DW-stochastic}), hence,  the mean time to switch (the mean first passage time) becomes longer and longer.  In order to make a reliable switch we need to minimise this error by making sure the noise is optimised for a typical applied bias.  On the other hand, if the noise is too low, then for a physical range of bias forces and fixed sampling time, the device may not change its state at all. Hence, there is a trade-off between switching rate and error, and as we will see this trade-off works its way through the entire learning process.

By coarse graining over the noise in the two meta-stable states of the switch, we see that in this case the activation switch  implements a binary switch with values $\pm 1$. While the average behaviour is a real variable described by  a smoothly varying activation function, the actual value taken by the physical switch in each trial is a binary variable. 

\subsection{Thermodynamics of a classical activation switch}
\label{activation-switch}
The double-well model is a good example in which to discuss the thermodynamics of an activation switch. The switching of the system between the two metastable states is caused by a static bias force internal to the system.  Modulating the external bias as a function of time changes the slope of this bias force just enough to cause a switch from the left to the right with high probability over the time that the bias acts.

Suppose we begin with the particle localised on the left, so the output of the switch is $n=-1$.  If the bias is low enough it will stay in this state for a very long time: it is a meta-stable state. As the bias force is increased in time, work is being done on the particle. This continues until the noise causes a transition to the right hand side and the value of the switch changes to $+1$.  This means the time taken to switch, and the work done in each trial, is a random variable. Once the device crosses the barrier it must dissipate energy (as heat in this case) through friction. Clearly the heat dissipated is also a random variable and varies from one trial to the next just as the work done varies form one trial to the next.  What is the probability distribution for these two random variables? To answer this question we need to turn to the field of stochastic thermodynamics and fluctuation theorems~\cite{Seifert, Halpern_2015}.

Suppose we first focus on the steady states before and after the bias is changed. The change in the thermodynamic state is given in terms of the change in the Helmholtz free energy $\Delta F=\Delta E-T\Delta S$ where $E$ is the internal energy, $S$ is the system entropy and $T$ is the temperature of the heat bath. The change in free energy is calculated with respect to the two thermal steady states before and after the bias is changed.

In our model, the bias force is a time-dependent parameter, $\lambda(t)$, that takes the system from one meta-stable steady state to another. As $\lambda(t)$ is changing  the energy of the particle is changing as work is being done on the system. Buy observing the system over a single trial, we eventually see a switching event, after some time interval, $\tau$, which is a random variable.  Rather than directly finding the probability distribution for $\tau$, and the work done for that trial, we consider ensemble averages of these quantities over many trials. Remarkably, we can find an equality between these ensemble averages and the change in thermodynamical free energy between initial and final steady states. 

One such relation is given by Jaryznski's theorem~\cite{Jarzynski} that states
\begin{equation}
    {\cal E}[e^{-\beta w}]  =e^{-\beta \Delta F},
\end{equation}
where the average on the left hand side is the ensemble average over many trials with a time varying bias force,  $w$ is the work done in each trial, and $\beta=1/k_BT$.  As has often been noted, this is remarkable as it enables one to  experimentally determine a thermal equilibrium quantity from a non-thermal equilibrium process. In the case of an activation switch, work is done {\em on} the system by the bias forces, thus the change in free energy is positive $\Delta F >0$ no matter which transition ($\pm 1\longrightarrow \mp 1$) takes place.

The noise in this model is thermal, and switching takes place by thermal activation over a barrier. As the noise is reduced by reducing temperature, switching will take longer and longer to realise. What happens as the temperature is reduced to zero?  Of course temperature cannot be reduced to zero. What we actually mean is that thermal energies are small compared to the quantum of action.  As the system enters the quantum domain switching occurs through quantum tunnelling, leading us to introduce a {\em quantum activation switch} and the corresponding quantum learning machines that these can implement.  

There have been many studies of thermal fluctuation theorems for biased double-well potentials. The case of a double well with linear bias that is varied linearlly in time is discussed in~\cite{NickelsenEngel2012}. Rather than continue to use the double-well model for an activation switch we will turn to a two-state stochastic model that is more generally applicable and easily generalised to the quantum case. 

\subsection{Two state activation switches.}
\label{binary-SA}
The double well model can be coarse-grained in terms of a two state Markov model~\cite{Gardiner} for a binary  state labelled $n\in\{-1,1\}$ with energies $E_{-}, E_+$ where $E_{+}-E_-=E_0$.  For convenience we use $E_{\pm}=\pm E_0/2$. The response to time-dependent external forces is described by an in-homogeneous Markov process~\cite{Rao}
\begin{equation}
\label{classical-me}
    \frac{dp_1(t)}{dt} =\mu p_{-1}(t)-\nu p_1(t) =-\frac{dp_{-1}(t)}{dt}, 
\end{equation}
where $p_n(t)$ is the probability of finding the system in state $n$ at time $t$ and
$\mu(t)(\nu(t))$ is the rate of transition from $n=-1(n=1)$ to $n=1(n=-1)$.
In the absence of time dependent bias forces the transition rates are time independent, $\mu(t)=\mu, \nu(t)=\nu$, and   the steady state probabilities of the two level system are
\begin{eqnarray}
\label{ss}
p_{-1} & = & \frac{\nu}{\mu+\nu},\\
p_1 & = & \frac{\mu}{\mu+\nu},
\end{eqnarray}
which follows from detailed balance. The mean energy is given by $\bar{E}=E_0 (p_1-p_{-1})/2$. 

 In the classical over-damped double-swell case, the rates depend on the details of the potential and the temperature of the environment. For example, if the initial state is in thermal equilibrium at temperature $T$, then 
$
p_1/p_{-1} = e^{-\beta E_0}
$
where $\beta =(k_B T)^{-1}$, and so the initial rates satisfy
$
\mu/\nu =e^{-\beta E_0}
$
by detailed balance. In the case of a negative bias force, Kramer's formula~\cite{Gardiner} gives
$
    E_0 = V(x_*)-V(y_*)
$
where $x_{*}<0, y_*>0$ are the stable fixed points of the dynamics. For example, if $V(x)=x^2(x^2-1)-\lambda x$, then $E_0=0$ at zero bias, $\lambda=0$. For $|\lambda| << 1 $ we find $E_0$ increases quadratically around $\lambda=0$.     
The average value of $n$ is 
$
    \bar{n}=\tanh(\beta E_0)
$
and 
\begin{equation}
    p_1=\frac{1}{2}(1+\tanh(\beta E_0)).
\end{equation}  

On the other hand if an external system maintains the system in a negative temperature state~\cite{negative-temperature} then
$
p_1/p_{-1} = e^{\beta E_0}.
$
Hence, again by detailed balance
$
\mu/\nu =e^{\beta E_0}.
$
More generally, we do not need to assume any relation between rates and temperature and thus will not further specify the stochastic process beyond the Markov transition rates in anticipation of our treatment of quantum activation switches.

\begin{figure}
    \centering
    \includegraphics[scale=0.5]{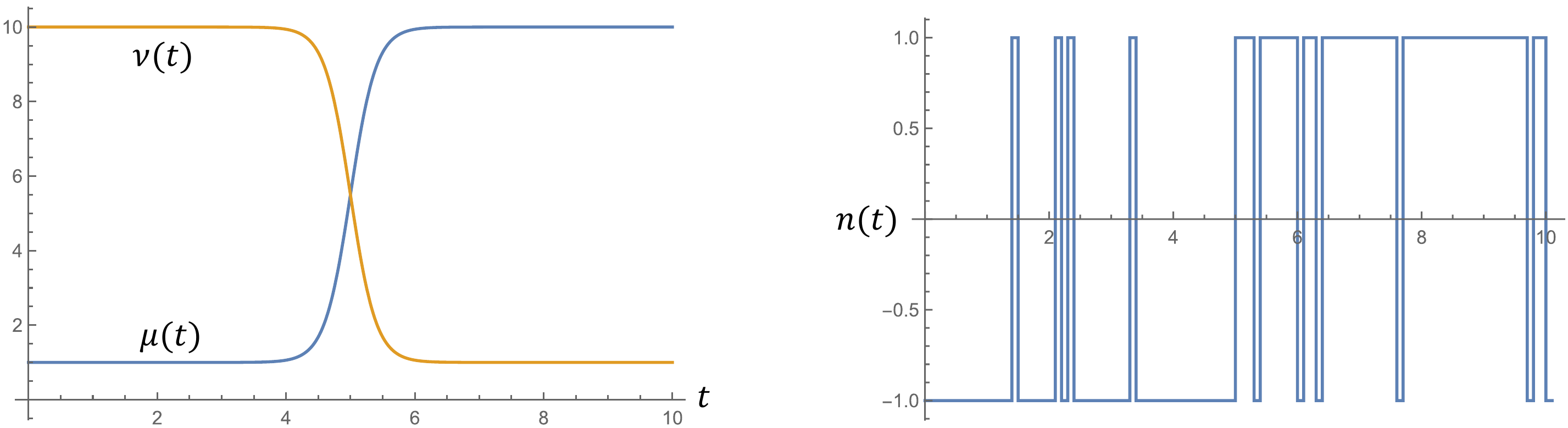}
    \caption{An example of an in-homogeneous Poisson process as defined in Eq. (\ref{rates}) with $f(t)$ defined as a sigmoid function centred on $t_0=5.0$ and $\mu=1.0$, $\nu=10.0$. In the left we see that the transition rates swap at time centred on $t_0=5$. On the right we see a simulation of the observed process as a function of time. }
    \label{sigmoid-switch}
\end{figure}

When a time dependent bias force acts, rates become time dependent. For example, 
\begin{eqnarray}
 \label{rates}
 \mu(t) & = & \mu +f(t)(\nu-\mu), \\\nonumber
 \nu(t) & = & \nu -f(t)(\nu-\mu),
\end{eqnarray}
where $f(t)$ varies smoothly from a minimum of zero at $t=0$ to a maximum of $1$ at $t=\tau$. This ensures that at the end of time variation we have a simple swap of the rates $\mu\leftrightarrow \nu$. As an example we will take the piece-wise continuous function
\begin{equation}
    f(t)=\left \{\begin{array}{cc}
            \frac{t}{\tau} & 0\leq t\leq \tau,\\
            0 & \mbox{otherwise}.
            \end{array} \right .
\end{equation}
Another example based on a sigmoidal switching function is shown in Fig.~(\ref{sigmoid-switch}).

If we assume that, in the absence of the perturbation, the system is in the steady state $n=\pm 1$, with the probability given in Eq. (\ref{ss}), then the probability of the system being in the state $n=\pm 1$ at time $t=\tau$ is given by 
\begin{eqnarray}
p_{-1}(\tau) & = & \frac{\mu}{\mu+\nu}+\frac{(\nu-\mu)}{\tau(\nu+\mu)^2} \left (1-e^{-(\mu+\nu)\tau}\right ),\\
p_1(\tau) & = & \frac{\nu}{\mu+\nu}-\frac{(\nu-\mu)}{\tau(\nu+\mu)^2} \left (1-e^{-(\mu+\nu)\tau}\right ).
\end{eqnarray}
In the limit that $(\mu+\nu)\tau >> 1$, the initial distribution is swapped and the entropy is unchanged, but in a general in-homogeneous process the entropy would increase. In the same limit the change in the average energy of the system is 
\begin{equation}
\label{average-energy}
    \overline{\Delta E} =E_0\left (\frac{\nu-\mu}{\nu+\mu}\right ).
\end{equation}

In a single system view, we can define a classical stochastic process $n(t)\in\{-1,1\}$ that obeys the classical stochastic differential equation
\begin{equation}
    dn(t) =\frac{1}{2}(1-n(t) )dN_+(t)-\frac{1}{2}(1+n(t) )dN_-(t),
\end{equation}
where $dN_\pm(t)$ are  Poisson processes defined by
\begin{eqnarray}
{\cal E}[dN_-(t)] & =&  \nu(t)dt,\\
{\cal E}[dN_+(t)] & =&  \mu(t)dt.
\end{eqnarray}
The first term is zero when $n(t)=+1$ indicative of the fact that the system cannot make an upward transition in energy from this state. The second term is zero when $n(t)=-1$, indicative of the fact that the system cannot make a downward transition in energy from this state. The measured value of the energy is defined by $E(t)=E_0(1+n(t))/2$.

We can now consider a simplified version of the thermodynamic fluctuation theorems.
 Regardless of the value of $n$ at the start of the action, the probability that it does not change is   
\begin{equation}
    Pr(s=0) =e^{-\int_0^\tau\ dt (\mu(t)+\nu(t))}=e^{-(\nu+\mu)\tau}\equiv \eta.
\end{equation}
Note that this goes to zero as $(\mu+\nu)\tau\cal\rightarrow \infty$.
The probability for each of the other two cases are given by 
\begin{eqnarray}
Pr(s=1) & = & \frac{\mu}{\nu+\mu}(1-\eta)+\left (\frac{\nu-\mu}{\mu+\nu}\right ) \left (1-\frac{(1-\eta)}{\tau(\mu+\nu)}\right ),\\
Pr(s=-1) & = & \frac{\nu}{\nu+\mu}(1-\eta)-\left (\frac{\nu-\mu}{\mu+\nu}\right ) \left (1-\frac{(1-\eta)}{\tau(\mu+\nu)}\right ).
\end{eqnarray}

In the limit that $(\mu+\nu)\tau\rightarrow \infty$, the average value of $w$ over many trials is then seen to be 
\begin{equation}
\label{two-state-ft}
    {\cal E}(w)=E_0\left (\frac{\nu-\mu}{\nu+\mu}\right ).
\end{equation}
Using Eq.~(\ref{average-energy}) we now see that $ {\cal E}(w)=\overline{\Delta E}$. This is an example of a statistical fluctuation theorem for systems not necessarily in thermal contact with a heat bath. 
If  $\nu >> \mu$, the average work is positive and work is done on the system. If  $\mu>>\nu$ (the negative temperature case), the average work is negative and the system does work. As we have assumed that $(\mu+\nu)\tau>> 1$, the entropy does not change, and the change in the Helmholtz free energy is equal to the change in the mean energy. 

\begin{figure}
    \centering
    \includegraphics[scale=0.5]{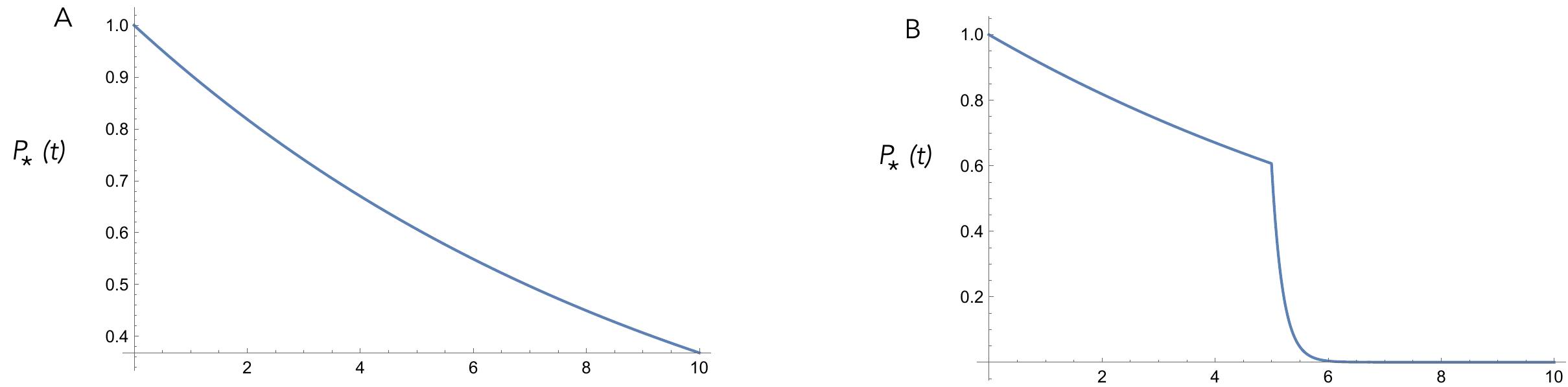}
    \caption{Two examples of the wait time distribution for a switch starting in the state $1$. In (A) the transition rate out of the initial state, $n=1$, is a constant $\nu=0.1$. In (B) the transition rate is impulsively increased at $t=5$.  }
    \label{wait-time}
\end{figure}

The time a system takes to change its state is the {\em first passage time}. In Fig.~(\ref{wait-time}), we plot the first passage time distributions for the case $\mu > \nu$, initially prepared in the state $1$. 
The transition rate out of this state is given by $\nu(t)=\nu(1+a f(t))$.  When $a=0$, the rate is constant and we are dealing with a pure Poisson process. In that case, the wait time distribution is simply exponential as is clear in Fig.~(\ref{wait-time}A). If the rate is time dependent, the wait time distribution is 
\begin{equation}
    P_*(t)=\exp[-\int_0^t dt' \nu(t')].
\end{equation}
As an example we take $f(t)=\theta(t-t_0)$ where $\theta(x)$ is the unit step function. We see that when the modulation of the transition switches on, the system is very likely to make an immediate transition to the state $0$, Fig.~(\ref{wait-time}B).

\subsection{A partially observed activation switch}
\label{partial-observe}
The previous discussion was based on a `single-shot' view of an activation switch: a single control pulse changes the transition rates and the binary output $n\in\{-1,+1\}$ responds accordingly. We will need to consider situations for which we do not have direct access to the output value of the switch; the activation switch may trigger a change in real valued physical quantity which is directly measured, e.g., the voltage across a circuit element. In this case, the output may be subject to measurement noise in addition to the noise associated with the switching probability. 

Let the physical quantity be represented by $x\in{\mathbb R}$. We define the response of this variable to the output $n$ by a conditional probability $P(x|n)$. The unconditional statistics (i.e. we do not know the actual value of $n$) of the output variable is then given by 
\begin{equation}
    P(x,t)=\sum_{n=-1}^1 P(x|n)p_n(t).
\end{equation}
We may also calculate the \emph{conditional} state of the switch given a particular observation $x$. This is given by Bayes rule as 
\begin{equation}
  p_n(t|x) = \frac{P(x|n)p_n(t)}{P(x,t)}.
\end{equation}
As an example, consider the Gaussian
\begin{equation}
    P(x|n) =(2\pi\Delta)^{-1/2} e^{-(x-\chi n)^2/2\Delta},
\end{equation}
where $\chi$ is a constant, with the same units as $x$, which defines the separation of the two gaussian peaks corresponding to $n=\pm 1$ in the distribution of $P(x)$ , and $\Delta$ defines the width of each peak. While $\chi$ depends on how strongly the switch is coupled to the physical quantity we measure, $\Delta$ depends on the number of measurements we perform to sample $x$, hence, in the case of a very poor measurement, $\Delta >>\chi^2$. It is easy to see that the mean and variance of the observed  process  are given by
\begin{eqnarray}
\bar{x}  & = &  \chi(p_1-p_{-1})=\chi\bar{n},\\
{\cal V}[x] &= & \Delta +4 \chi^2 p_1 p_{-1}.
\end{eqnarray}
We see the noise is additive: the first term shows the measurement noise and the second term is proportional to the noise associated with the switching probability by noticing that the variance in $n$ is simply $4p_1p_{-1}$. 
We can define a time continuous version of this measurement by making a sequence of very many such measurements repeated  $N>>1$ times in a time interval $[t, t+\delta t)$ such that $\mu\delta t, \nu\delta t <<1$. Let us now form the effective time average 
\begin{equation}
    y(t) =\frac{1}{N}\sum_{j=1}^N x_j.
\end{equation}
The law of large numbers then implies that the mean and variance of $y(t)$ are given by  
\begin{eqnarray}
\bar{y}(t) & = & \chi \bar{n},\\
{\cal V}[y(t)] & = & \frac{\Delta}{N}.
\end{eqnarray}
We introduce the rate of measurement as $\gamma=N/\delta t$ such that $\sqrt{\Delta}=  \gamma \delta t $, and further define $\chi =\kappa \delta t$. The continuous limit is now taken by assuming that as $N=\gamma \delta t \rightarrow \infty $, the ratio $\chi/\sqrt{\Delta} =\kappa/\gamma$ is fixed. This fixes the ratio of the peak separation to the peak width of the two gaussian peaks in the distribution for $P(x)$ as the continuous limit is taken. Then, we see that  ${\cal V}[y(t)] \propto \delta t$. This is a diffusion process, so we can write
\begin{equation}
y(t) = \kappa \bar{n} dt+\sqrt{\gamma}dW.
 \end{equation}
If the system is in one state or the other at time $t$, the observed process is subject to diffusion at rate $\gamma$. We write this in terms of a stochastic current $y(t) = I(t) dt$. As no physical device can respond infinitely fast to the white noise process we assume that the actual observed current is filtered according to 
\begin{equation}
    I_o(t)= r\int_{-\infty}^t e^{-r (t-t')}I(t') dt',
\end{equation}
where the exponential defines the response function of the filter and $r$ is the response rate.
Hence, the observed current obeys the stochastic differential equation
\begin{equation}
    dI_o(t) = -r^2 I_o(t)dt + r\kappa \bar{n} dt+r\sqrt{\gamma}dW.
\end{equation}

\begin{figure}
    \centering
    \includegraphics[scale=0.75]{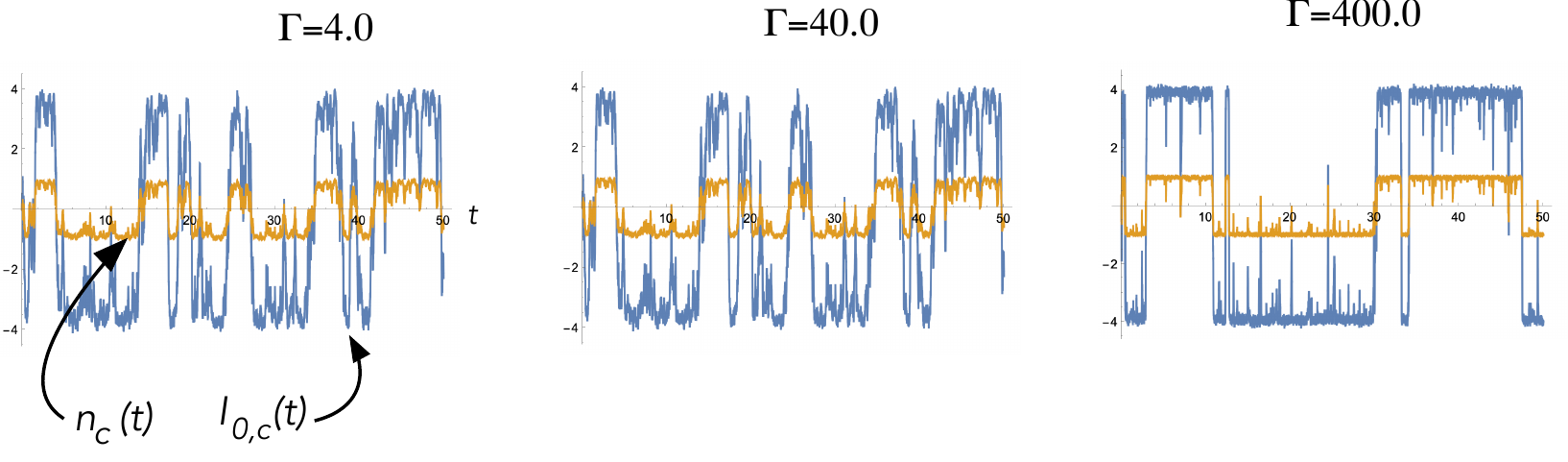}
    \caption{ Samples of the conditional observed process, $I_{O,c}(t)$, and the conditional mean, $n_c(t)$, as a function of time for three different values of $\Gamma$. Parameters are $r=20.0, \kappa=40.0$, and $\mu=\nu=0.5$.}
    \label{observed-trial}
\end{figure}

In a similar way we can compute the conditional state of the switch given a particular result $y$ in a time $\delta t$. Let $p_{c,n}$ be the conditional state of the process given an entire history of results $y(t)$ up to time $t$.  Using Baye's theorem we see that
\begin{equation}
p_{c,n|y}=\frac{P(y|n)p_{c,n}}{P(y)}.
\end{equation}
After a little algebra, we find that 
\begin{eqnarray}
   p_{c, 1|y} & = & \frac{p_{c,1}}{p_{c,1}+p_{c,-1}e^{-2N\chi y/\Delta}},\\
    p_{c, -1|y} & = & \frac{p_{c,-1}}{p_{c,-1}+p_{c,1}e^{2N\chi y/\Delta}}.
\end{eqnarray}
Using the scaling with $\delta t$ we find that the arguments of the exponential are $\pm 2\kappa y/\gamma$. As $y$ is of order $\delta t$, we can expand the exponential to lowest order in $\delta t$ taking care that as $dW^2$ is of the order of $\delta t$, we should include the Ito correction. If we calculate the conditional mean by using $\bar{n}_c= p_{c, 1|y}-p_{c, -1|y}$, we find the conditional increment of the mean over a time step $dt$ is 
\begin{equation}
    d\bar{n}_c(t)= -\sqrt{\frac{\Gamma}{2}}(1-\bar{n}_c(t)^2) dW,
\end{equation}
in which $\Gamma = 2\kappa^2/\gamma$, and
where we have used $p_{c,\pm 1}= (1\pm \bar{n}_c(t))/2$. Including the background stochastic switching, the conditional equation of motion becomes
\begin{equation}
\label{conditional-mean}
    d\bar{n}_c(t)= \mu(1-\bar{n}_c(t))dt/2-\nu (1+\bar{n}_c(t))dt/2 -\sqrt{\Gamma/2}(1-\bar{n}_c(t)^2) dW,
\end{equation}
and the corresponding observed process is 
\begin{equation}
\label{observed-processss}
    dI_{o,c}(t) = -r^2 I_{o,c}(t)dt + r\kappa \bar{n}_c dt+r\kappa\sqrt{2/\Gamma}dW. 
\end{equation}
A strong measurement corresponds to $\Gamma >> \kappa$. In this limit, the noise on the observed process is small. 
In Fig.~(\ref{observed-trial}), we plot a sample of the observed current, given by Eq.~(\ref{observed-processss}), as a function of time for increasing values of $\Gamma$.

  \section{Quantum activation switches}
  
In the previous section, we did not need to assume that the initial and final distributions on the interval $[0,\tau)$ were thermal equilibrium distributions.  This enables us to describe a quantum activation switch operating at close to zero temperature where quantum noise dominates thermal noise.  

  \subsection{Quantum dots}
  \label{Quantum-Dots}
\begin{figure}
    \centering
    \includegraphics[scale=0.4]{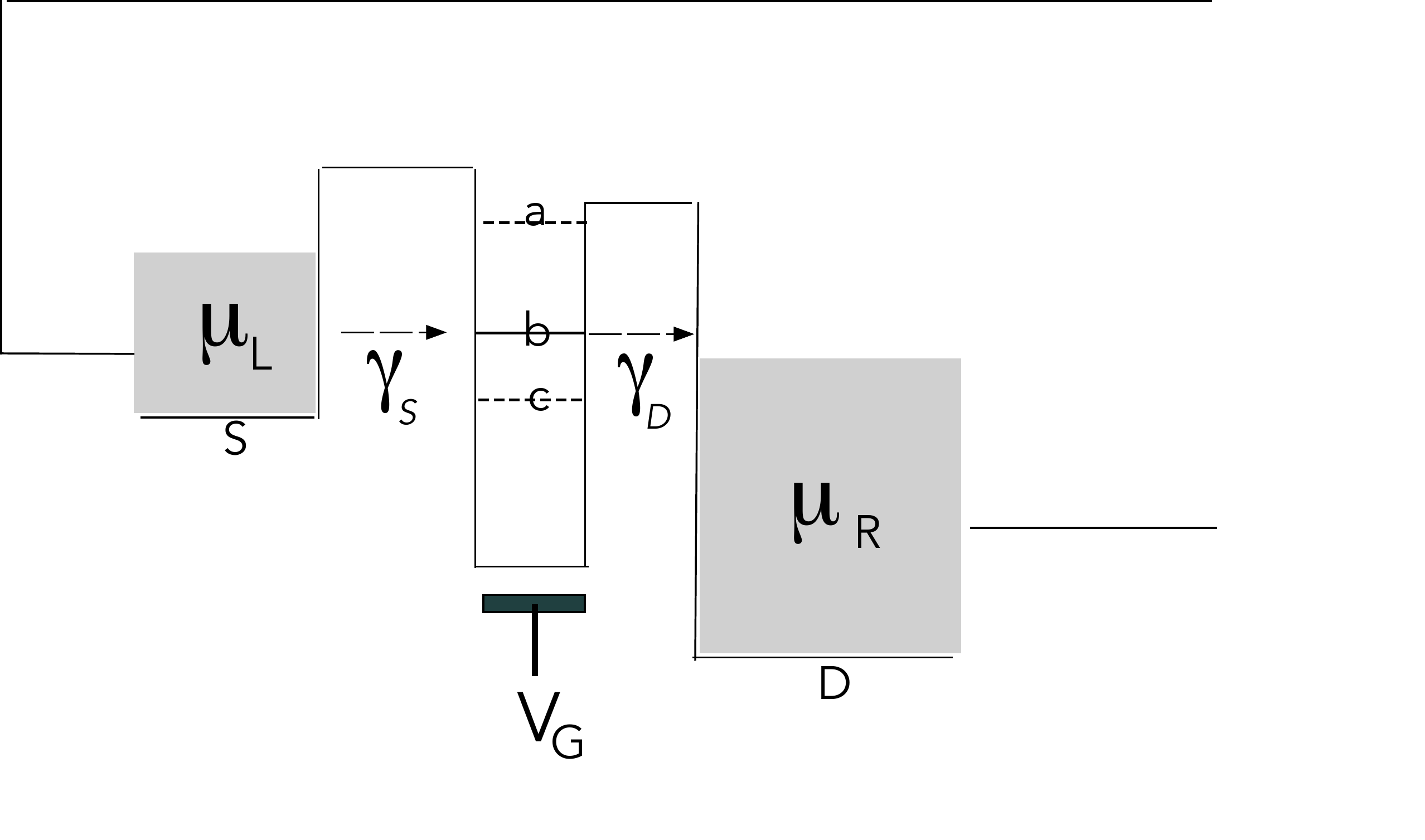}
    \caption{A nanoelectronic quantum dot. Electrons can tunnel from the Fermi reservoir, the source(S), onto the dot and then from the dot to the Fermi reservoir, the drain(D). The bare rates of tunnelling through the left barrier is $\gamma_D$, while the rate of tunnelling through the right barrier is $\gamma_S$. The source drain bias voltage $V_{SD}$ ensures a current flows provided the dot is not blockaded (a,c). A gate voltage changes the energy of the dot with respect to the source and drain enabling current to flow in the case of (b) but not (a) and (c). The current is thus controlled by both $V_{SD}$ and $V_G$. The difference in the chemical potentials is $\mu_S-\mu_D=V_{SD}$. }
    \label{QD}
\end{figure}

The simplest example of such a quantum activation switch comes form the field of nanoelectronics~\cite{Heikkila}. In one kind of such a device, a single electron can tunnel through a barrier from one Fermi reservoir, the source, into a quasi bound state before tunnelling out into a second Fermi reservoir, the drain. This is shown schematically in Fig.~(\ref{QD}).
Each Fermi reservoir is maintained in local thermodynamics equilibrium by an external circuit. When an electron tunnels from the source to the dot, an increment of charge is drawn into the source from the external circuit to restore the chemical potential $\mu_L$. On the other hand, if an electron tunnels off the dot into the drain, an increment of charge is given up to the external  circuit to maintain the chemical potential $\mu_R$. In this way, fluctuations in the occupation of the dot determine current fluctuations in the external circuit. This current is measured in the experiment.  

\begin{figure}
    \centering
    \includegraphics[scale=0.75]{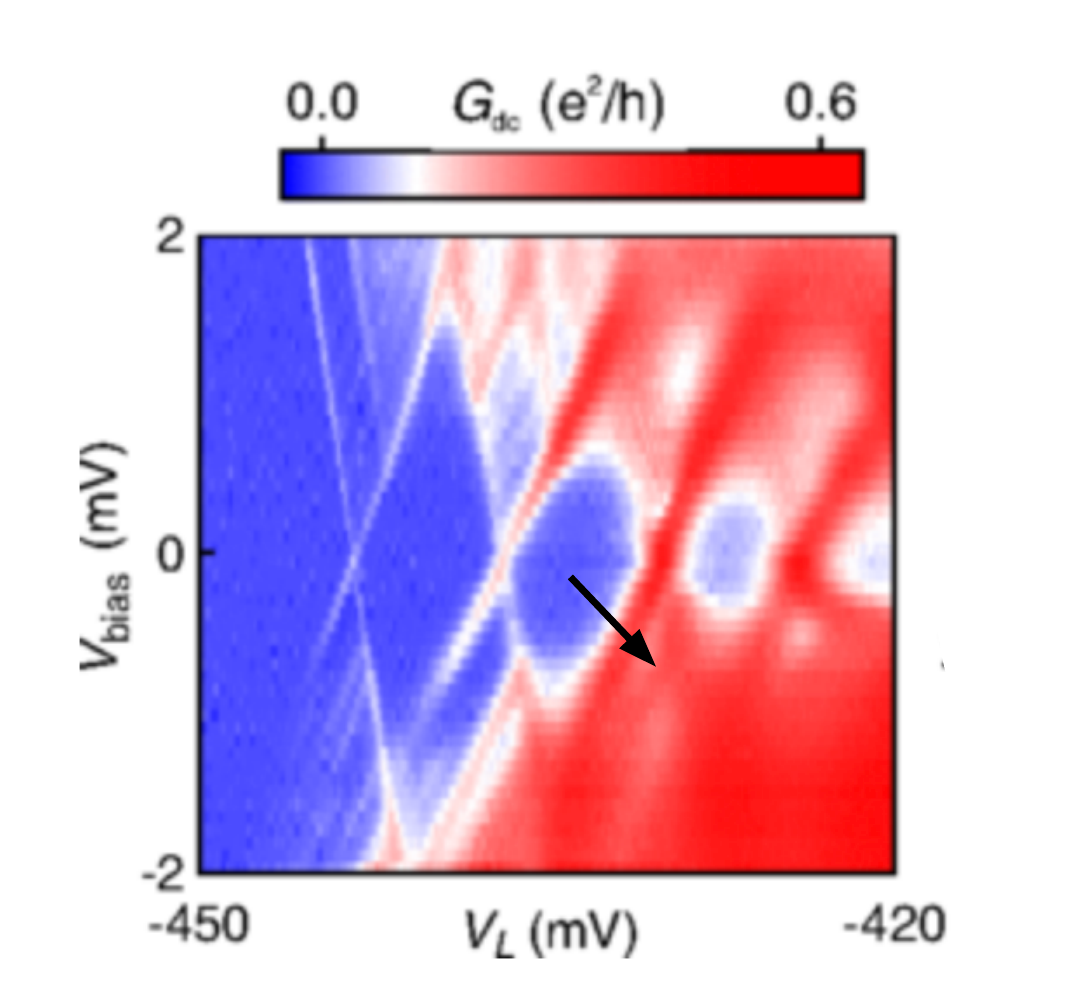}
    \caption{Coulomb diamonds: conductance is plotted in  grey scale versus the source drain voltage, $V_{bias}$, and the tunnel gate voltage, $V_L$.  Current flows when the conductance is high. The arrow indicates a bias change that causes the switch to transition from low  to high occupation of the dot. (Plot courtesy of Natalia Ares, Oxford. ) }
    \label{QD-CB}
\end{figure}

In a {\em Coulomb blockade} regime (see Fig.~(\ref{QD})), current cannot flow as there is no energy available in the source (a) or no state available in the drain (c). The rate of tunnelling through the left barrier is $\gamma_S$ and the rate of tunnelling though the right barrier is $\gamma_D$. These rates are a function of the barrier widths and often controlled with additional voltage gates.  If the gate voltage is adjusted, tunnelling through the dot is possible and current flows as electrons tunnel onto the dot at rate $\Gamma_{in}=\gamma_Sf_S+\gamma_D f_D$  and off at rate $\Gamma_{out}=\gamma_S(1-f_S)+ \gamma_D(1-f_D)$. Here, $f_S$ and $f_D$ are, respectively, the Fermi factors in the source and the drain. They vary between $0$ and $1$ and depend on the temperature and the local chemical potential in each reservoir. At very low temperatures, $f_S\approx 1$ and $f_D\approx 0$, so $\Gamma_{in}=\gamma_S$, and $\Gamma_{out}=\gamma_D $. In summary, there are two voltage controls, $V_{SD}$ and $ V_G$, to set the condition for a current to flow. This is shown in terms of the Coulomb blockade diamond diagram, Fig.~(\ref{QD-CB}).

The two control voltages can be ramped independently around some fixed position giving time dependent tunnelling rates that can vary from zero to gigahertz rates. This is modelled using the Markov process
\begin{equation}
  \frac{dp_1}{dt}= \Gamma_{in}(1-p_1)-\Gamma_{out}p_1,  
\end{equation}
where $p_1(t)$ is the probablity that the dot is occupied. 
The change in the average occupation of the dot is given by
\begin{equation}
   \frac{dn(t)}{dt}=\Gamma_{in}(1-n(t))-\Gamma_{out}n(t).
\end{equation}
The steady state occupation of the dot is 
\begin{equation}
    \bar{n}_{ss} = \frac{\Gamma_{in}}{\Gamma_{in}+\Gamma_{out}}.
\end{equation}

\begin{figure}
    \centering
    \includegraphics[scale=0.4]{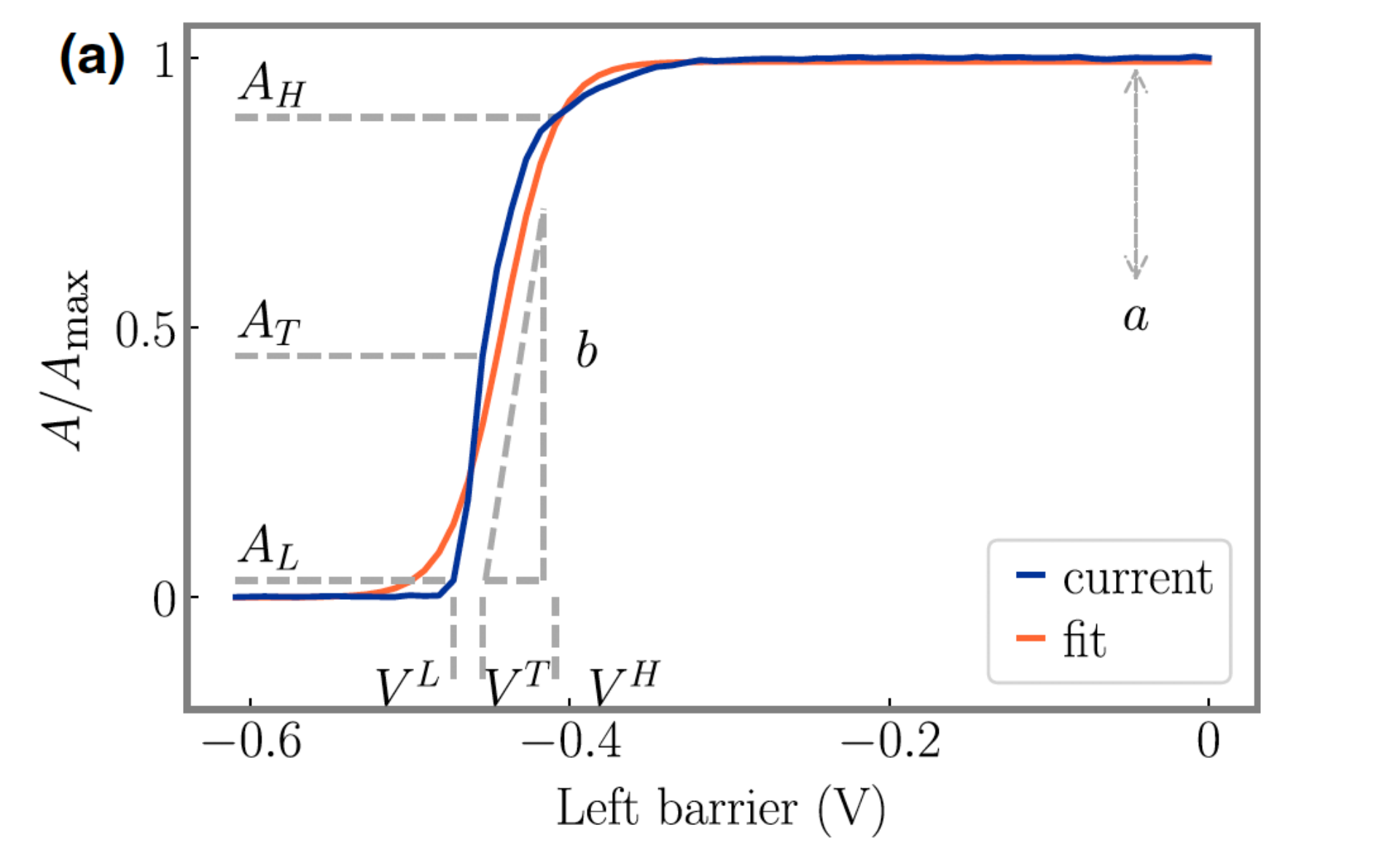}
    \caption{The average current versus a gate voltage as a gate changes the tunnel barrier between the source and the dot. Also, shown is a best-fit curve as discussed in the text. Adapted from~\cite{Darulova}. 
}
    \label{darulova}
\end{figure}

As an example, we take the surface-gate defined quantum dot device described in Darulov\'{a} et al.~\cite{Darulova}. In this experiment, we are interested in the average current that flows as the gate controlling the left barrier is varied. A typical result is shown in Fig.~(\ref{darulova}). 
The average current varies from off to on as the tunnel barrier width is decreased by changing the voltage, $V$, of a surface gate. The voltages $V^L,V^H$ and $V^T$ refer to cutoff, transition and saturation gate voltages. Also shown is a best-fit curve 
\begin{equation}
    f(x,a,b,c)=a[1+\tanh(bx+c)],
\end{equation}
as discussed in~\cite{Darulova}, where  
\begin{equation}
    x=\frac{v-v_{min}}{v_{max}-v_{min}},
\end{equation}
in which $v_{min}$ and $v_{max}$ are the lowest and highest voltage set points respectively,  $a$ is the amplitude of the current variation above and below the transition current, $A_T$, $b$ is the slope of the curve at the transition voltage $V_T$, and $c$ is chosen such that $f=0$ at $v=v_{min}$.  This curve is a typical average response for an activation switch as previously discussed and thus looks promising for implementing a learning machine primitive element. 

In terms of the inhomogenous Markov process discussed in section (\ref{binary-SA}), such an average response function follows from setting
\begin{eqnarray}
    \mu(x) & = & \gamma(1+\tanh (bx+c))/2,\\
    \nu(x) & = &  \gamma.
\end{eqnarray}
The average current thus varies from $0$ to the steady state current, $A_{max}$. 
Operating as an activation switch means that the variable $x$ is a product of the bias weights and input and is effectively time dependent. The role of the parameter $b$ is discussed in section (\ref{classical-learning}).  We can thus adopt the results of the general case in section (\ref{partial-observe}). At very low temperatures, the Poisson rates are determined almost entirely by quantum tunnelling rates, so devices like this can function as true quantum activation switches. 

\begin{figure}
    \centering
    \includegraphics[scale=0.5]{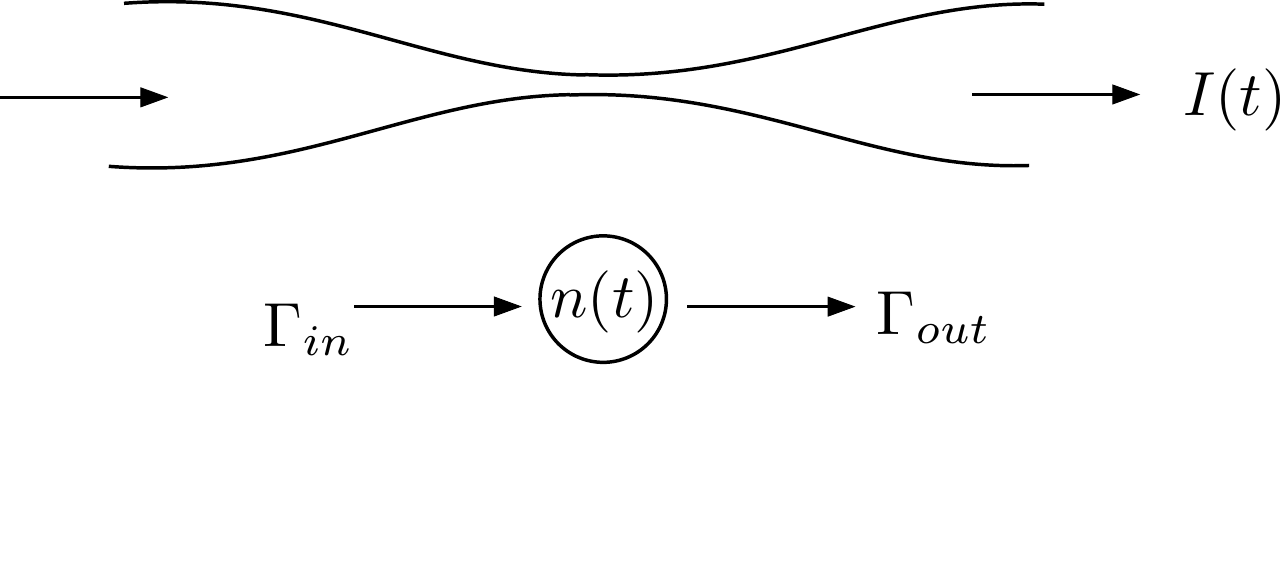}
    \caption{The current through a QPC can be controlled by electrostatic coupling to a nearby quantum dot.  If the occupation of the dot is described by quantum tunnelling point processes, the current through the QPC defines a continuous partially observed Markov process.   }
    \label{QPC-QD}
\end{figure}

We now need to consider what measurement process will be used to monitor the state of the quantum dot. The model presented in~\cite{Goan} uses a quantum point contact (QPC) and we will adopt that here. This is a measurement continuous in time. The idea is that the charge on the dot, $n(t)$, can apply a small conductance change to a narrow channel thus modulating the current, $I(t)$, through the channel. This is illustrated in Fig.~(\ref{QPC-QD}).

The measurement result is the current $I(t)$ and this is conditioned on the quantum state of the dot. Likewise, the measurement history conditiones the state of the dot.  These two effects are tied together, self-consistently, by a conditional Schro\"{o}dinger equation for continuous weak measurement~\cite{Goan}. We operate in a regime in which individual electron tunnelling events through the QPC cannot be resolved by the circuit of the QPC, and we can replace jump processes in the QPC by a continuous diffusive variable satisfying a Gaussian white noise distribution as in the case of the partially observed process in section (\ref{partial-observe}). However, the stochastic occupation of the dot modulates the current through the QPC.  The state of the dot is defined by a \emph{conditional} (inferred) occupation number $s_c(t)$. Note that this is not restricted to a binary number, but tends to that as the quality of measurement becomes very large (large $\chi$).  This leads to two Ito stochastic differential equations
\begin{eqnarray}
    ds_c(t) & = & (\Gamma_{in}(1-s_c)-\Gamma_{out} s_c)dt -2\chi s_c(1-s_c)dW,\\
    I_c(t)dt & = & \eta(1-2\epsilon s_c(t))+\sqrt{\eta} dW,
    \label{stochastic-dot}
\end{eqnarray}
where $\chi$ is a measure of the strength of the coupling between the dot and the conductance of the QPC channel, $\eta$ is an efficiency factor, and $\epsilon$ is a measure of how much the channel can be constricted when the dot is occupied. If we set $\epsilon=1$, the variation in the current is a maximum. The quantum-jump regime requires that $\chi$ is large. In a real experiment, we do not have access to $I(t)$ directly as the circuit has  finite response function. In effect, we can only observe $I(t)$ after a low pass filter. This is a quantum version of the partially observed process discussed in section (\ref{partial-observe}) with $s_c=(n_c+1)/2$ and $\kappa$ replaced by $\epsilon\eta$ determining the strength of the Coulomb coupling between the dot and the QPC and $\gamma=\eta$.

In the case of gate defined quantum dots, the tunnel barriers are also controlled by gate voltages~\cite{gate-QD}. There are many different kinds of gate defined quantum dots, so to be specific, we will focus on surface gate semiconductor devices.  In such devices, we can control the intrinsic tunnelling rates, $\Gamma_{in}$ and $\Gamma_{out}$, using applied voltages. For an activation switch it is sufficient to control $\Gamma_{in}$.

\section{Learning machines}
We are interested in  the thermodynamic limits of physical learning machines, be they quantum or classical. Before discussing particular schemes, it is important to understand that learning machines of any kind are necessarily dissipative devices.

\begin{figure}
\centering
 \includegraphics[scale=0.35]{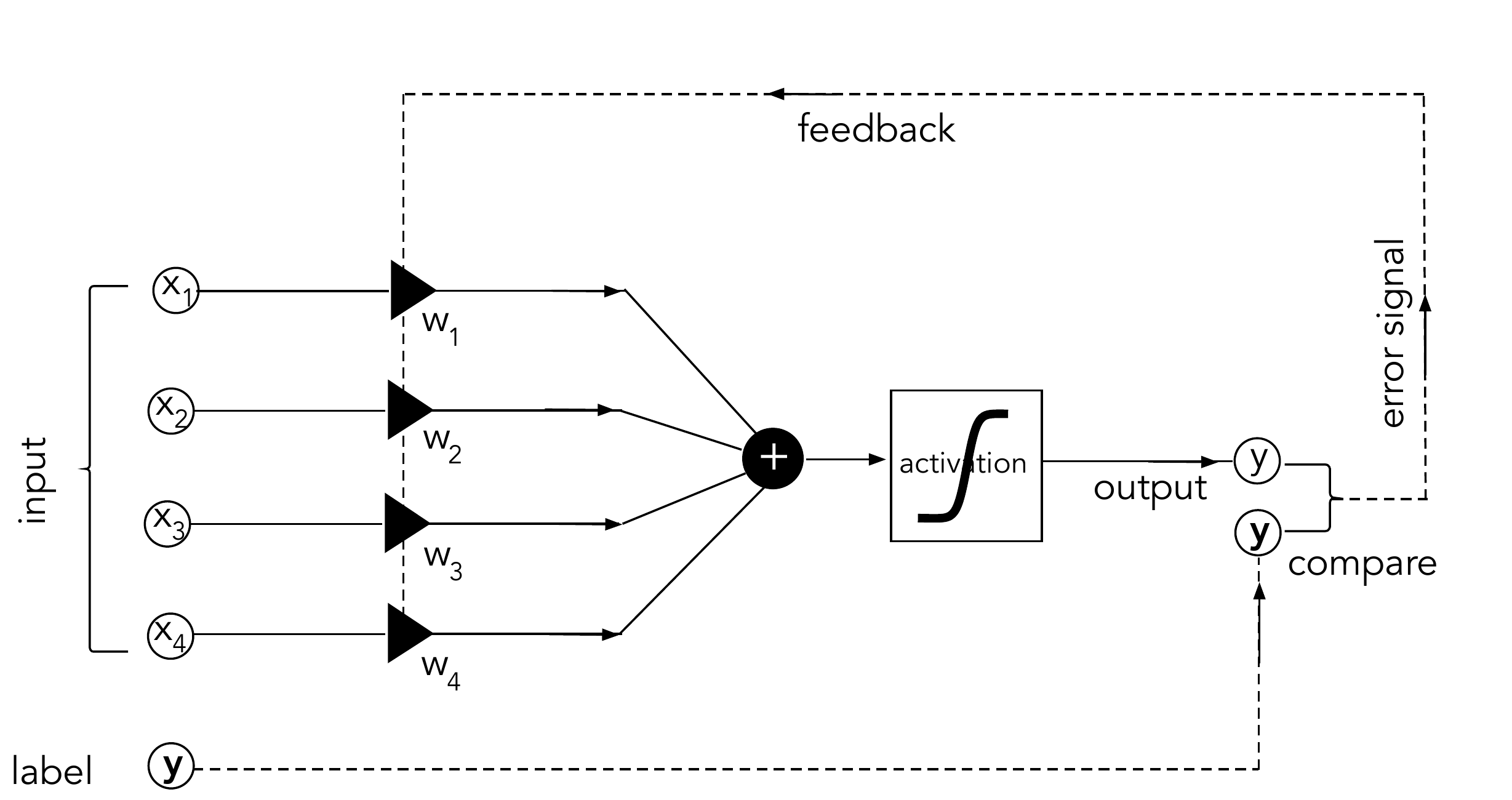}
 \caption{Learning based on a perceptron algorithm. The activation function, $\theta(z)$, is a binary-valued nonlinear function of a single real variable. Now define $\vec{x}$ and $\vec{w}$ as column vectors of the data variables, $x_j$, and the weights $w_j$.  Overall, the perceptron is a set of binary valued non linear functions, $f_{\vec{w}}(\vec{x})$, of the data, $\vec{x}$, indexed by the weights, $\vec{w}$. The algorithm proceeds in the following steps.  In a training mode, the data are labelled with a binary number ${\bf y}$, called the true label.  We then regard the output $y=f_{\vec{w}}(\vec{x})$ as a trial label for that data element. If $y={\bf y}$ on a single trial we do nothing, but if $y \neq {\bf y}$, we record an error and feedback to change the weights for the next trial of training data. The feedback is chosen to minimise the probability of an error in the long run. Once trained the perceptron can be used to assign a label to an unlabelled data element. }
 \label{perceptron}
\end{figure}

  As a simple example we will consider a perceptron model based on taking  the activation function to be the the Heaviside step function, $\theta(z)$, that takes value zero if $z<0$ and unity if $z\geq 0$. In fact, almost any non linear function will suffice. The learning algorithm is illustrated in Fig.~(\ref{perceptron}).

The perceptron algorithm will find a nonlinear function of many variables, the components of the data element weighted by corresponding weight vector $\vec{w}$, and correctly assigns a label to that data element, with low probability of error. Clearly this function is many-to-one and thus is not logically reversible, that is to say, we cannot obtain the input from the output. This is the entire point of learning: to reduce the correlations inherent in a large number of variables and represent them by a single function.  

\begin{figure}
    \centering
    \includegraphics[scale=0.25]{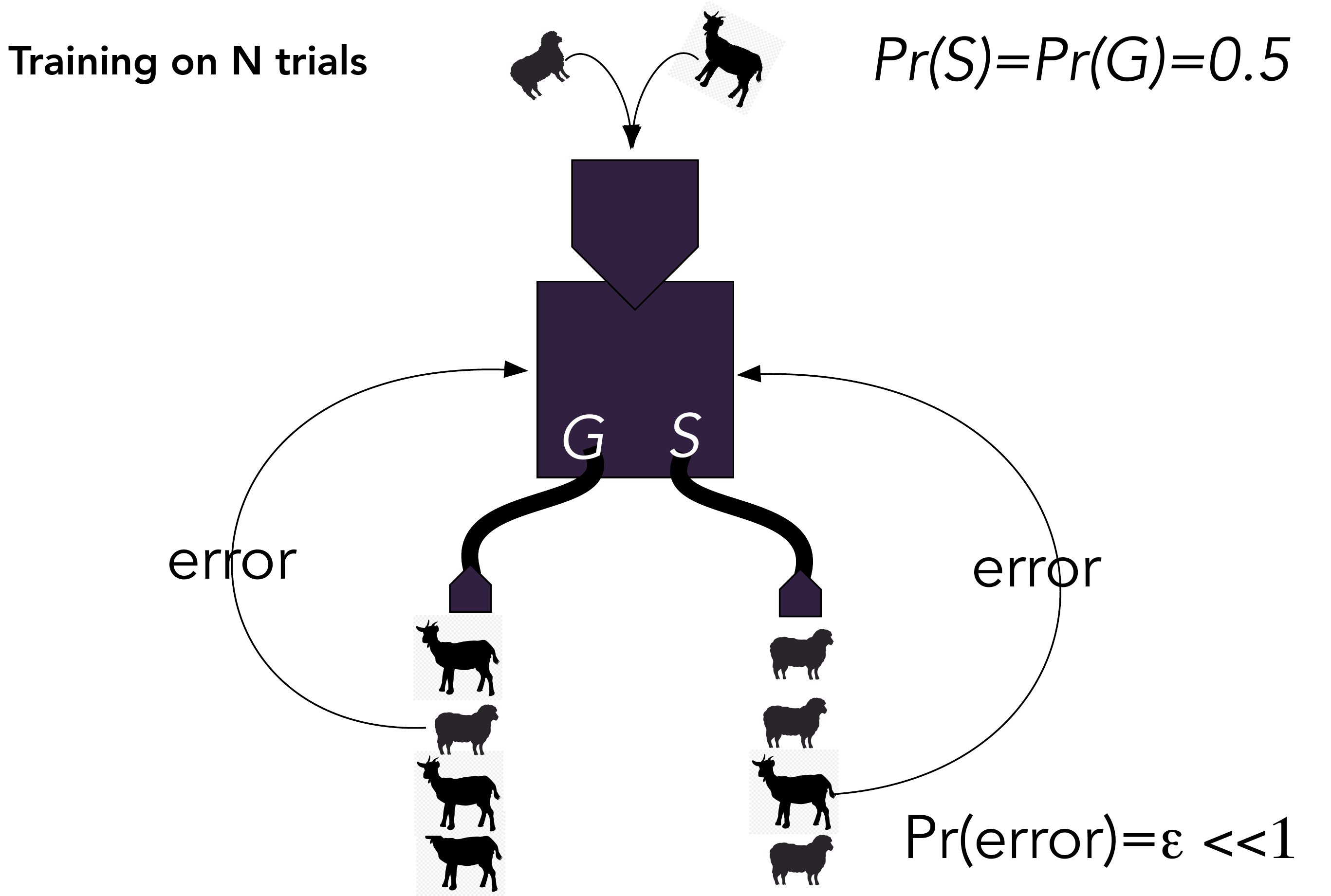}
    \caption{A learning machine is trained by feedback to classify images of goats and sheep. A correctly labelled prediction means the input image, chosen at random, comes out in the correct output.  At the end of a successful training run, the probability of error, that is the probability that an image comes out in the wrong channel, is very small $Pr(error) =\epsilon <<1$. This is achieved by actively feeding back to the machine parameters every time a mistake is made. The entropy at the input is $N\ln 2$ in natural units, where $N$ is the number of training examples.  Initially, the output has the same entropy but at the end of training, the entropy of the output records is reduced to roughly $N\epsilon$.  This must be paid for by an overall increase in the entropy of the machine's environment through heat dissipation that arises every time work is done in the feedback steps. }
    \label{sheep-goats}
\end{figure}

There is a deep connection between logical reversibility of an algorithm and physical reversibility of the hardware used to carry out the algorithm~\cite{Bennett,Bennett2003}.  Simply put, a logically irreversible function necessarily erases information from input to output. Landauer's principle~\cite{landauer1961irreversibility} connects this to thermodynamics. It says that there is a thermodynamic cost when information is erased.  It follows that any physical device that implements a logically irreversible function must pay a price in heat dissipated, and entropy generated,  in its environment.  We give a simple illustration of the necessary irreversibility of a machine implementing a perceptron algorithm in Fig.~(\ref{sheep-goats}).

While a learning algorithm, when implemented in hardware,  necessarily generates heat, it is far from clear what the minimum possible thermodynamic cost should be. Certainly current deep learning neural networks generate a lot of heat and consume vast amounts of energy~\cite{DNN-energy}. However, in recent years it has gradually been realised that this is unnecessary~\cite{McMahon, NTT}. One can build special purpose learning machines that operate far more efficiently from a thermodynamic cost perspective. What are the physical principles behind this?

\subsection{Classical learning machines}
\label{classical-learning}
In order to understand the thermodynamic costs of learning machines, in section \ref{LM-ML} we introduced the concept of an {\em activation switch}. This is an irreversible dynamical system  driven by bias forces --- the weights --- with a response that is a stochastic non linear function of its initial conditions --- the inputs. How can this be used to implement a physical learning machine? This will introduce an essential new element --- noise --- without which learning machines cannot operate. The fluctuation-dissipation theorem tells us that noise necessarily accompanies any irreversible process. This is another indication that learning and irreversibility are linked. The key difference between activation functions and activation switches is that the former is a function ${\mathbb R}\rightarrow {\mathbb R}$, the latter is described by a stochastic function  ${\mathbb R}\rightarrow n\in \{-1,+1\}$. It is the average of $x$ over many trials that is given by a real valued function of the weights and inputs.  

As an example we will use the in-homogeneous two state markov process described in section (\ref{binary-SA}) as an activation switch and build a perceptron model around this. This kind of model has been analysed in detail by Goldt and Seifert~\cite{GS}, including a careful thermodynamic study. We will first adapt their treatment to a discrete time protocol. We will make the physical feedback protocol as close as possible to the usual algorithmic learning process. The key difference is that in the physical device, the output of the activation switch is a random variable restricted to the values $n=\pm 1$, while in the algorithmic case, the output of the activation function is  is a real variable. The two become equivalent if we base the learning feedback on an estimate of the average output $\bar{n},$ over multiple trials using the same training data in each trial. This average is a real number.  


Each data point consists of an N -dimensional vector $\vec{\xi}$ and a label $n_T$ with 
\begin{equation}
\vec{\xi}=( \xi_1, \xi_2, \dots, \xi_N)^T,
\end{equation}
where $\xi_k\in \{-1,+1\}$. The label for each data vector is defined by $ n_T=\pm 1$. The objective is to train a perceptron to correctly identify the chosen feature of a randomly chosen data vector. A single training point is of the form $(\vec{\xi},n_T)$. 


The training of the perceptron is done by changing the transition rates between states but the bias forces now depend on a weighted sum of the components of a training vector.  If the two states represent a coarse-graining of an underlying double well potential,  the transition probabilities reflect thermal activation over a barrier depending on the bias forces applied~\cite{McW}. We will assume that the dynamical properties of the switch are such that whenever the bias changes the transition probabilities, it rapidly relaxes to the new steady state probabilities given by 
\begin{equation}
\label{two-state-ss}
    \frac{p_1}{p_{-1}} =\frac{\mu}{\nu},
\end{equation}
where $p_1$ is the probability that the switch is ON, and $p_{-1}=1-p_1$ is the probability that it is OFF. These are functions of the bias forces at each trial. The weights vector takes the form
\begin{equation}
\vec{w}=(w_1,w_2,\ldots,w_N)^T,
\end{equation} 
with components $w_k\in \mathbb{R}$.

The activation switch is so constructed that the probability for it to switch ON in a given trial, $p_1(\vec{w})$,  is a nonlinear  function of
\begin{equation}
\label{fire}
A(\vec{w})=\vec{\xi}\cdot\vec{w} =\sum_{i=1}^N \xi_i w_i+b,
\end{equation}
where $b$ is a bias. The probability for it to switch OFF in a given trial  is $p_{-1}(\vec{w})=1-p_1(\vec{w})$.  Note that we are thinking of the probabilities  as a function of the weights as that is what we will vary in training. 


The output of the activation switch in a single trial is defined as
\begin{equation}
n=\left \{\begin{array}{cc}
        +1 & \mbox{ON},\\
			-1 & \mbox{OFF}.
\end{array}\right .
\end{equation}
In order to set up a learning feedback process, we need a cost function to indicate how well the device is performing. This should be zero, or close to zero, when averaged over many trials when the learning is complete.   We define the cost variable at a single switching event as 
\begin{equation}
\label{random-error}
\epsilon =\frac{1}{4}(n_T-n)^2=\frac{1}{2}(1-n_T n),
\end{equation}
 where the right hand side is valid as in this case $n^2=1$. Note that $\epsilon$ is a random variable. The average error is given by
\begin{equation}
\label{av-error-prob}
\bar{\epsilon}(\vec{w}) =\frac{1}{2}(1-n_T \bar{n}(\vec{w})),
\end{equation}
where
\begin{equation}
\label{mean-out}
\bar{n}(\vec{w})=p_1(\vec{w})-p_{-1}(\vec{w})=2p_1(\vec{w})-1.
\end{equation}
If $n_T=1$, then $\bar{\epsilon}= p_{-1}$, while if  $n_T=-1$, then $\bar{\epsilon}= p_1$. In other words, the average cost is the error probability in each case. For example, if $n_T=1$ we require $n=1$ at that trial for a correct response. This means the detector fires.  If it fails to fire, that is an error, and it does so with probability $p_{-1}$. 

As $\epsilon$ takes the values $0,1$ at a single trial, we see that average of the square is equal to the mean. Thus the variance is
\begin{equation}
{\cal V}[\epsilon] = \bar{\epsilon}(1-\bar{\epsilon}).
\end{equation}
This is a maximum at $\bar{\epsilon}=1/2$ which occurs when $p_1=1/2$ and falls to a small value  as training proceeds. 

The change in the average error probability due to a variation in the weights is given by
\begin{equation}
\Delta \bar{\epsilon} =\Delta \vec{w}\cdot \vec{\nabla}_w\ \bar{\epsilon}.
\end{equation}
We need this to decrease as much as possible per trial so we set
\begin{equation}
\Delta \vec{w}= -\eta \vec{\nabla}_{\vec{w}}\ \bar{\epsilon},
\end{equation} 
where $\eta$ is a positive scaling constant. 
Thus
\begin{equation}
\Delta \bar{\epsilon}=-\eta|\vec{\nabla}_{\vec{w}}\ \bar{\epsilon}|^2.
\end{equation}
Hence, the feedback rule is  
\begin{equation}
\Delta\vec{ w}=\eta n_T\vec{\nabla}_wp_1(\vec{w}).
\label{feedback-rule}
\end{equation}

We now need to make a choice for the function that determines $p_1(\vec{w})$. We will take the sigmoidal form
\begin{equation}
\label{sigmoid}
 p_1(\vec{w})= \frac{1}{1+e^{-\beta A(\vec{w})}},
\end{equation}
where $\beta$ is a constant fixed by experimental design. In a thermally activated device it would be $\beta^{-1}= k_BT$. In a quantum tunnelling device it is some function of tunnelling rates. 

\begin{figure}
    \centering
    \includegraphics[scale=0.5]{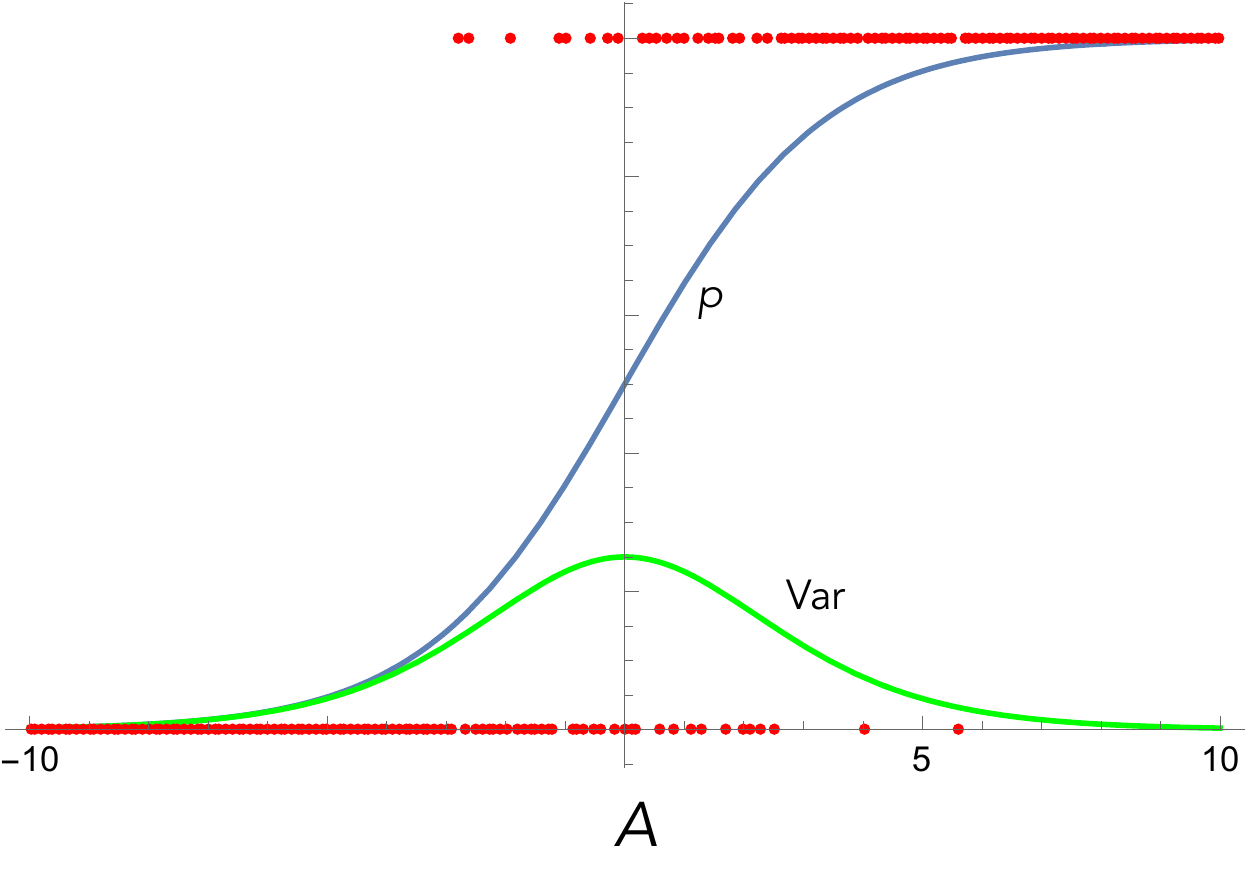}
    \caption{A plot of the probability for a Bernoulli trial for the switching of the activation switch as function of the activation variable $A$, as well as a corresponding random binary variable sampled at each point. Also shown is the variance of the distribution.  We set $\beta=0.6$. }
    \label{Bernoulli-example}
\end{figure}

In Fig.~(\ref{Bernoulli-example}), we plot the probability defined in Eq. (\ref{sigmoid}). This defines a Bernoulli distribution for the  `switch' events versus `no-switch'. The figure also shows a plot of the outcome of a random binary variable sampled from the distribution at each value of $A$. 
It is clear that at either extremes of the range the outcome is largely deterministic but around $A=0$ there are considerable fluctuations, as indicated by a peak in the variance. The range of this behaviour depends on $\beta$.

With the choice of Eq.(\ref{sigmoid}), we find that 
\begin{equation}
\vec{\nabla}_w\  p_1= \beta p_1(1-p_1)\vec{\xi}.
\end{equation}
Note that $p_1(1-p_1)$ is the variance of the Bernnouli distribution defined by $p_1$. If we use Eq. (\ref{mean-out}) to write this in terms of the mean $\bar{n}=2p_1-1$, then replace it in Eq. (\ref{feedback-rule}), we find the feedback rule as 
\begin{equation}
\label{delta-weight}
\Delta \vec{w}=\eta  \frac{n_T\beta(1-\bar{n}^2)}{4}\vec{\xi}.
\end{equation}
The change in the weights depends on both the  training data and the corresponding true labels. Note that this goes to zero as $\bar{n}\rightarrow \pm 1$, corresponding to learning the required function. With this choice
\begin{equation}
\Delta \bar{\epsilon}=-\eta\beta^2(1-\bar{n}^2)^2/16.
\end{equation} 
This is always negative, and has a maximum when $\bar{n}=0, (p_1=1/2)$. At the start of training this is the case and the change in the average error is large. As training proceeds it decreases. 

\begin{figure}
\centering
\includegraphics[scale=0.6]{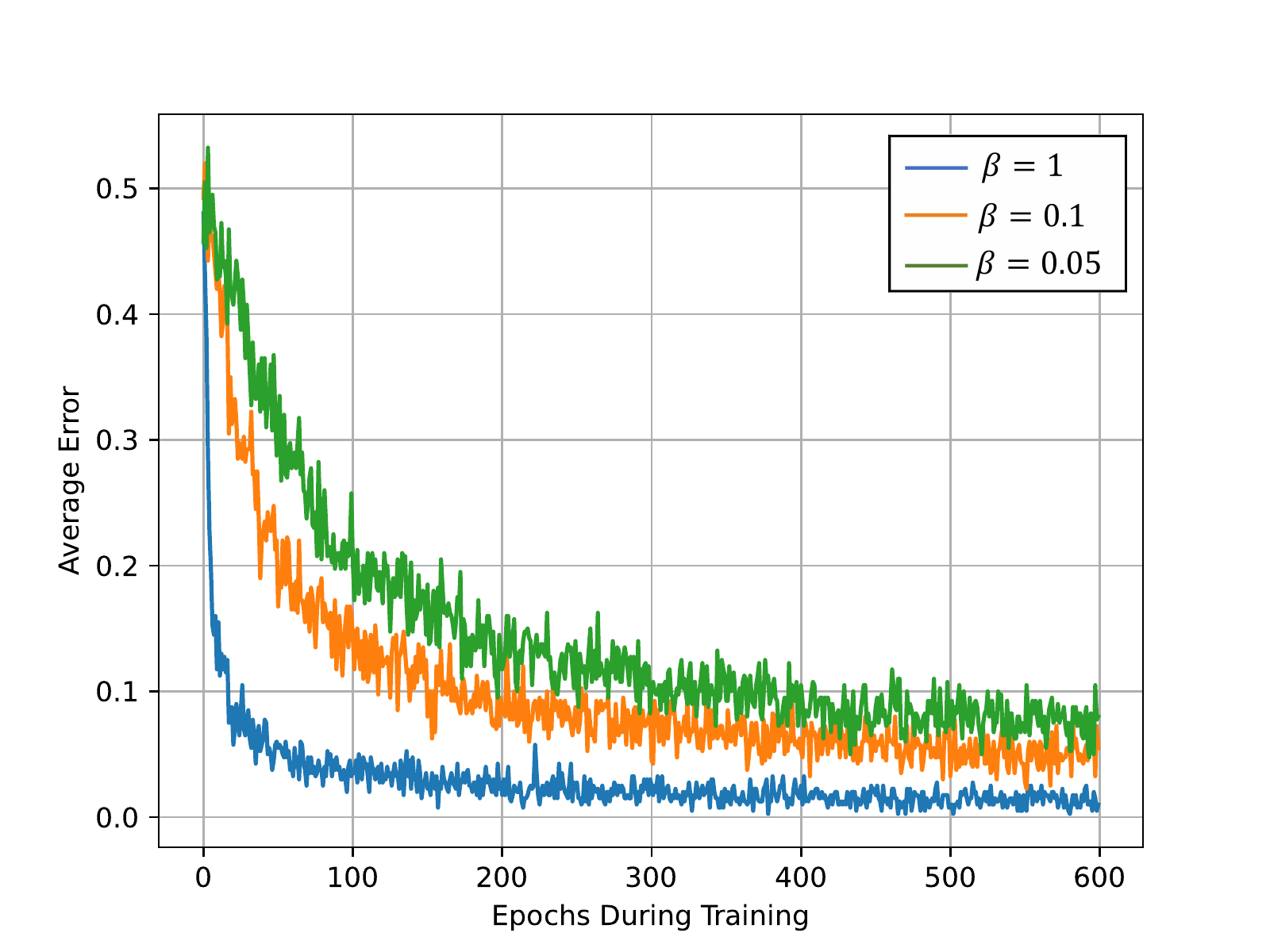}
\caption{ Plot of average error versus epoch number for different values of $\beta$ for learning NOT logic gate. The initial values of weight and bias are randomly chosen as $w=0.01$, $b=1$, the learning scale, $\eta=1$, and the number of samplings in each trial is 200. The values of weight and bias after training become $w=-4.43$, $b=0.01$ for $\beta=1$, $w=-2.99$, $b=0.01$ for $\beta=0.1$, and $w=-2.54$, $b=0.04$ for $\beta=0.05$.}
\label{NOT-sim}
\end{figure}

The training protocol requires that the weight vector, $\vec{w}$, be adjusted at each trial to minimise the average error probability, $\bar{\epsilon}$.  However, this means that we need to sample this distribution, equivalently we need to sample $\bar{n}$. In effect, this is simulating an individual experimental run of a physical machine, and averaging over a set of runs.   Each trial needs to be composed of many repetitions with the same input data, labels and weights. For example, consider learning NOT logic gate.  This can be done with a single perceptron.  The two correctly labelled inputs are
 $(x,n_T) \in\{(-1,1), (1,-1)\}$. In the first batch, or `epoch', we input, say, $100$ trials of  one of these chosen at random, say $(-1,1)$ and form the average of $n$ over these $100$ trials. Given this estimate for $\bar{n}$, we compute an estimate for $\bar{\epsilon}$ using
\begin{equation}
\label{error-prob}
\bar{\epsilon} =\frac{1}{2}(1-n_T \bar{n}).
\end{equation}
The weights are then adjusted according to Eq. (\ref{delta-weight}). A new input and label is chosen at random and the process repeated. 
In Fig.~(\ref{NOT-sim}), we show an example of learning NOT.  

Changing $\beta$ changes the noise associated with switching events. A very large value of $\beta$ means that switching rarely takes place except for values of $A$ near zero (see Fig(\ref{Bernoulli-example})). A very small value of $\beta$ means that switching takes place for a large range of values for $A$. Clearly the error rate falls more quickly as $\beta$ increases. We thus expect that fast learning requires a large value of $\beta$ and initial weights near zero. 

However, this is not the full story. The problem comes in determining how many trials, $N$, to include in each epoch to estimate $p(\vec{w})$. The uncertainty in the estimate of this probability is $\delta p$ and given by the  Cramer-Rao lower bound for the Bernoulli distribution~\cite{cover2012elements}
\begin{equation}
    \delta p^2 \geq \frac{p(1-p)}{N}.
\end{equation}
This bound is always worst near $p=1/2$, requiring a larger number of trials than at either extreme. Note estimating $p(\vec{w})$ is equivalent to estimating $\bar{n}$

\begin{figure}[!t]
\centering
\includegraphics[scale=0.7]{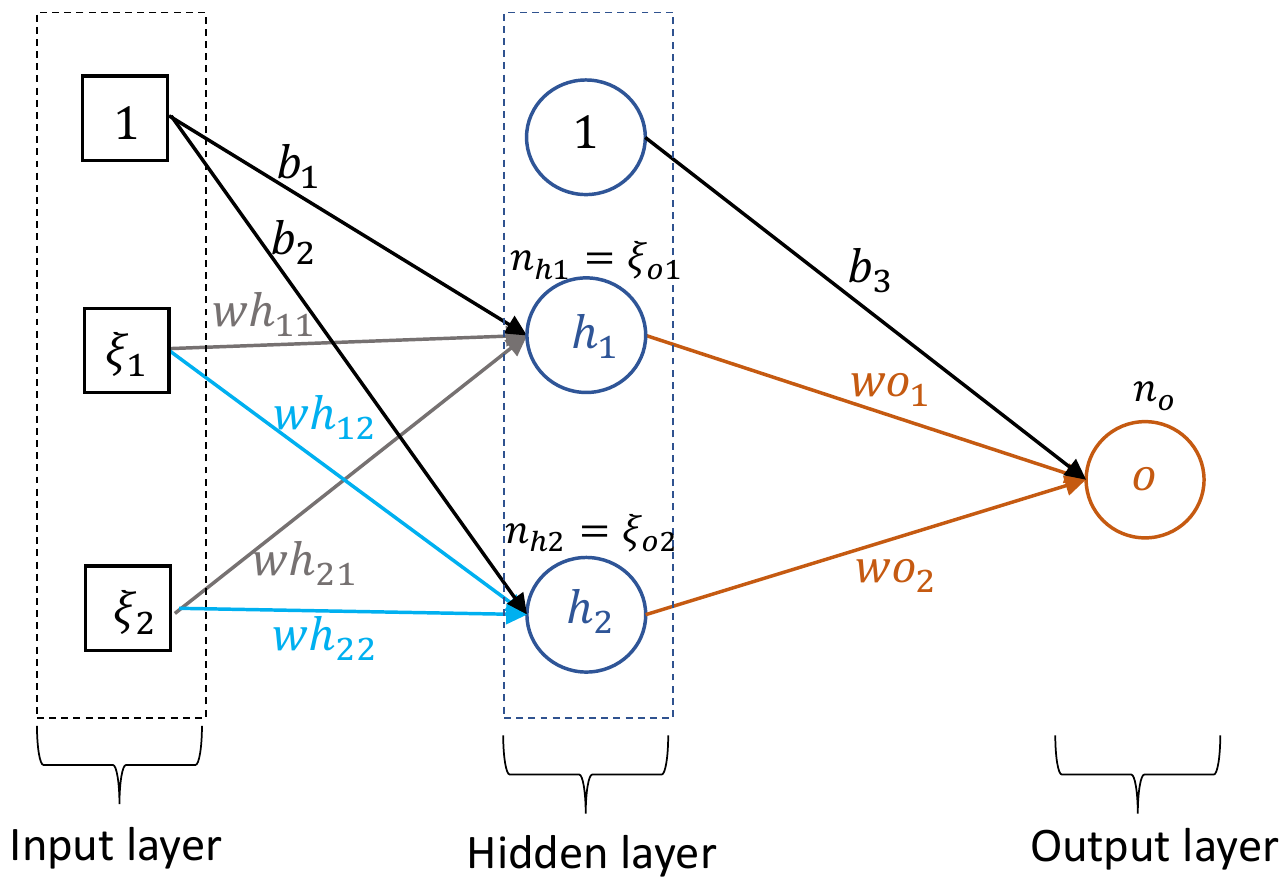}
\caption{A two layer neural network model to learn XOR logic gate. The input vector $\vec{\xi}$ has two components, $\xi_1$ and $\xi_2$. The hidden layer consists of two neurons, $h_1$ and $h_2$ and  the output layer has one neuron, $o$. The weights forwarded from the input layer to hidden layer are labeled by $wh_{ij}$, where $i (i=1,2)$ labels the component of the input layer and $j (j=1,2)$ labels the hidden neuron. The weights from the hidden neuron $j$ to output neuron are shown by $wo_j$. There biases are shown by $b_k$, where $k=1,2$ labels the biases for the hidden layer and $k=3$ refers to the output bias. The components of the input vector to the output neuron, $\xi_{oi} (i=1,2)$ are given by the output of the hidden layer such that we have $\xi_{oi}=n_{hi}$.}
\label{XOR-NN}
\end{figure}

In the above example, we showed the application of an activation switch as a linear classifier. However, by building a network of these activation switches we can mimic the mathematical machine learning algorithms used in deep learning neural networks to extend the application of the learning machines to solve nonlinear classification problems. To demonstrate the operation of the activation switches as nonlinear classifiers, we look at the simple example of learning XOR gate by using the same inhomogeneous two state Markov processes described in previous example. 
\begin{figure}
\centering
\includegraphics[scale=0.5]{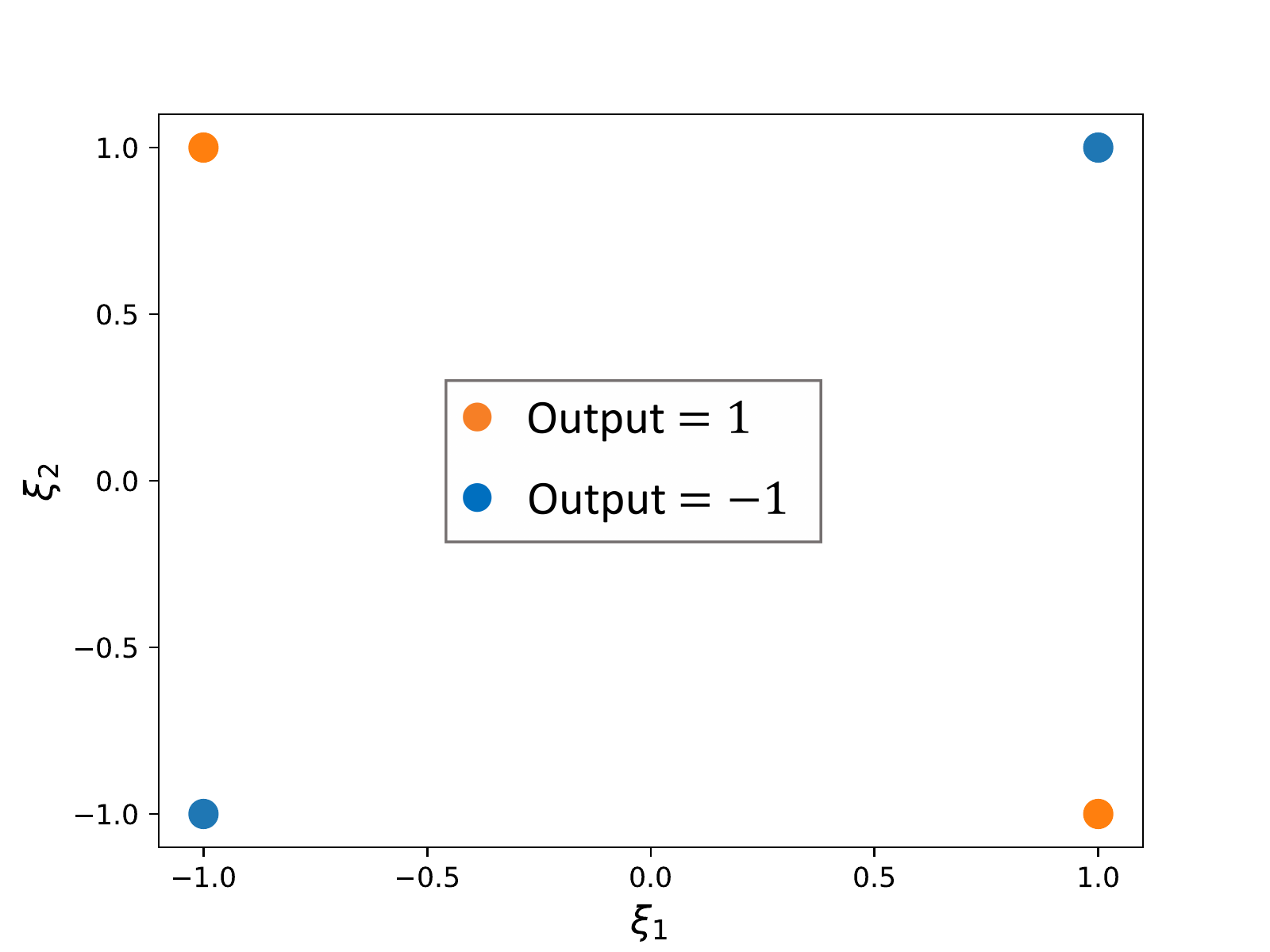}
\caption{Graphical representation of a logical XOR gate.}
\label{XOR-graphically}
\end{figure}
Fig.~(\ref{XOR-graphically}) shows graphical representation of XOR gate with components $1$ and $-1$. It is clear that in case of XOR we can not linearly classify the input data. It is known that a neural network consisting of a hidden layer with two neurons and an output layer with one neuron, as shown in Fig.~(\ref{XOR-NN}), can learn XOR logic gate. Here, we build the same network but replace the mathematical neurons with physical neurons. 
The four labeled inputs are 
\begin{equation}
([\xi_{1},\xi_{2}],n_{T})\in\{([-1,-1],-1),([-1,1],1),([1,-1],1),([1,1],-1)\}.\nonumber
\end{equation}
The cost variable at a single switching event is given by  
\begin{equation}
\label{random-error2}
\epsilon =\frac{1}{4}(n_T-n_o)^2=\frac{1}{2}(1-n_T n_o),
\end{equation}
in which $n_o$ is the final output of the network, see Fig.~(\ref{XOR-NN}). Hence, the average error in each epoch, made of sampling of the outputs for a given input over many trials, becomes
\begin{equation}
\label{error-mean}
\bar{\epsilon} =\frac{1}{2}(1-n_T \bar{n}_o)
=\frac{1}{2}(1+n_T)-n_Tp_1^{[o]},
\end{equation}
where $p_1^{[o]}$ is the probability of the output neuron switch ON and following the previous example for NOT gate, we take a sigmoid form given by 
\begin{equation}
 p_1^{[o]}= \frac{1}{1+e^{-\beta A_{o}}},\hspace{0.5cm}A_{o}=\vec{\xi}_o\cdot\vec{wo}+b_3 =\sum_{i=1}^2 n_{hi} wo_{i}+b_3,
\label{p1o}
\end{equation}
where the input to the output layer, $\vec{\xi}_o$ is the output of the hidden layer, $\vec{n}_h$. As shown in Fig.~(\ref{XOR-NN}), $\vec{wo}$ is the weight vector associated with the output layer with two elements $wo_i$, $i=1,2$, and $b_3$ is the output bias. In the forward pass, the probability that hidden neuron, $h_i$, switches ON is
\begin{equation}
 p_1^{[h_j]}= \frac{1}{1+e^{-\beta A_{hj}}},\hspace{0.5cm} A_{hj}=\vec{\xi}\cdot\vec{wh}+b_j =\sum_{i=1}^2 \xi_i wh_{ij}+b_j,
\label{p1h}
\end{equation}
where $\vec{wh}$ is the weight matrix associated with the hidden layer with four elements $wh_{ij}$, $i,j=1,2$, and $b_j$, $j=1,2$, is the bias associated with $h_j$. 
For a given input, $([\xi_{1},\xi_{2}],n_{T})$, we run the forward pass many times and sample the outputs of the hidden layer, $\vec{n}_h$, and the output layer, $n_o$. We then calculate the averages of these output values to run the backward pass. For that, we follow similar procedure as in previous example to find the feedback rules for the hidden and output weights and biases by using the gradient descent method. First, in order to decrease the average error as much as possible in each epoch we set
\begin{equation}
 \Delta \vec{wo}=-\eta \vec{\nabla}_{\vec{wo}}\ \bar{\epsilon}.
\label{}
\end{equation}
By using chain rule, the feedback rule for the output layer can be written as
\begin{eqnarray}
 \Delta \vec{wo}&=&-\eta \vec{\nabla}_{p_1^{[o]}}\ \nonumber \bar{\epsilon}\cdot\vec{\nabla}_{\vec{wo}}p_1^{[o]}\ \\\nonumber
 &=&\eta n_T\vec{\nabla}_{\vec{wo}}p_1^{[o]}\ \\
 &=&\frac{1}{4}\eta n_T\beta (1-\bar{n}_o^2)\bar{n}_{h},
\end{eqnarray}
where the hidden layer average output vector, $\bar{n}_{h}$, is a two dimensional vector such that $\bar{n}_{h}=(\bar{n}_{h_1},\bar{n}_{h_2})$. In a similar way, We find the feedback rule for hidden layer by moving the hidden weights and biases into the direction of the minimum error such that
\begin{equation}
 \Delta \vec{wh}=-\eta \vec{\nabla}_{\vec{wh}}\ \bar{\epsilon}.
\label{}
\end{equation}
Hence, the change in hidden weights is given by
\begin{equation}
 \Delta \vec{wh} = -\eta \vec{\nabla}_{p_1^{[o]}}\  \bar{\epsilon}\cdot\vec{\nabla}_{p_1^{[h]}}p_1^{[o]}\cdot\vec{\nabla}_{\vec{wh}}p_1^{[h]},
 \label{delta-w_h}
\end{equation}
where $p_1^{[h]}=[p_1^{[h_1]},p_1^{[h_2]}]$. We can calculate the terms of the above expression as
\begin{equation*}
\vec{\nabla}_{p_1^{[o]}}\  \bar{\epsilon}=-n_T,
\end{equation*}
\begin{equation*}
\vec{\nabla}_{p_1^{[h]}}p_1^{[o]}=2\beta p_1^{[o]}(1-p_1^{[o]})\vec{wo},
\end{equation*}
\begin{equation*}
\vec{\nabla}_{\vec{wh}}p_1^{[h]}=p_1^{[h]}\cdot(1-p_1^{[h]})\cdot\vec{\xi}.
\end{equation*}
Therefore, the feedback rule to update the hidden weights can be written as 
\begin{equation}
\Delta wh_{ij}=\frac{1}{8}\eta n_T\beta (1-\bar{n}_o^2)wo_j(1-\bar{n}_{hj}^2)\xi_i.
\end{equation}

\begin{figure}
\centering
\includegraphics[scale=0.6]{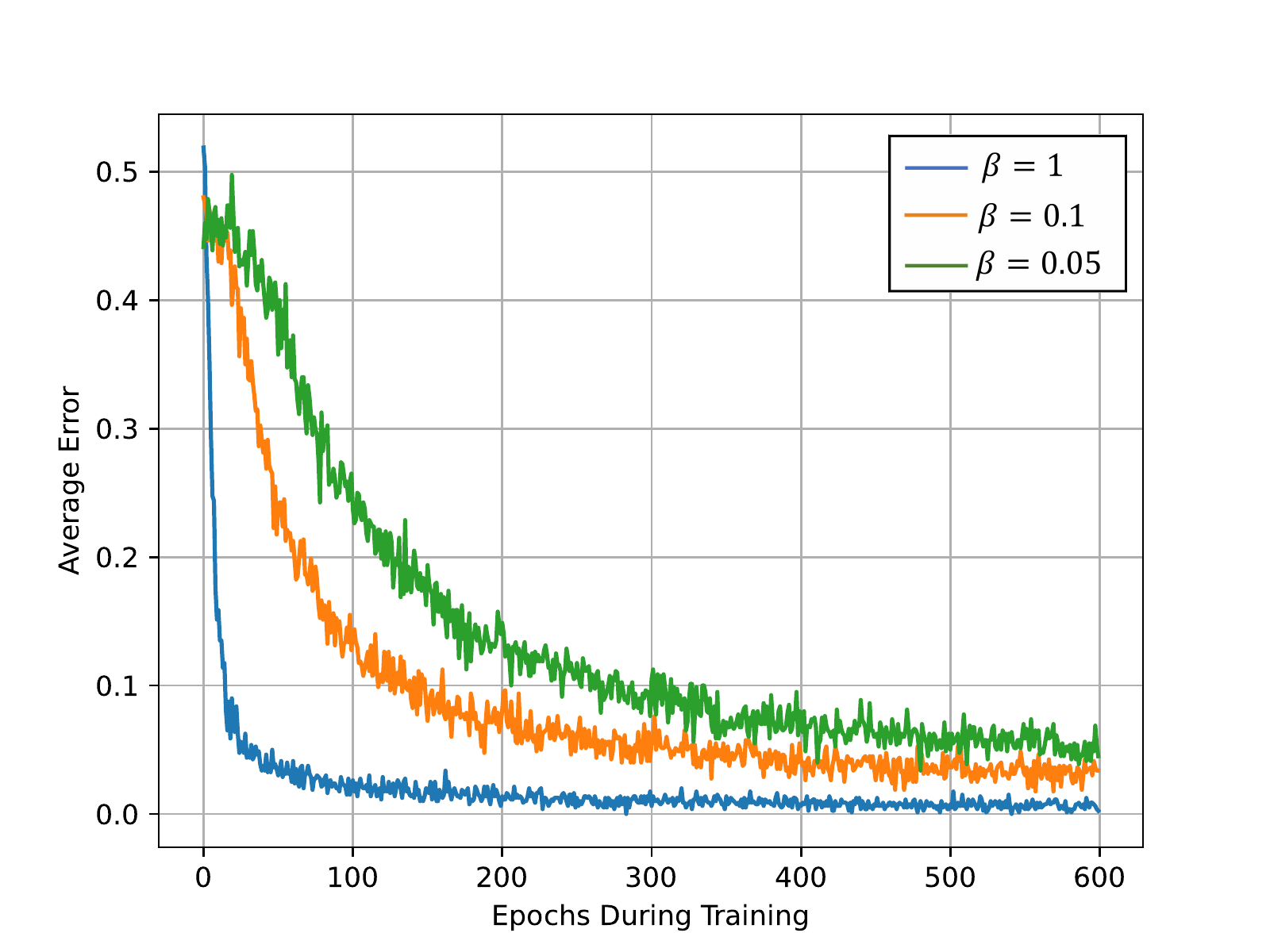}
\caption{Average error versus epoch number for different values of $\beta$ for learning XOR logic gate. The initial values of weights and biases for this plot are $\vec{wh}=\big(\begin{smallmatrix} -5 & 8\\ -2 & 3 \end{smallmatrix}\big)$, $\vec{wo}=\big(\begin{smallmatrix}  -2 \\ -3 \end{smallmatrix}\big)$, $b_h=\big(\begin{smallmatrix}  -1 \\ -3 \end{smallmatrix}\big)$, $b_3=-1$ and learning scale, $\eta=1$. Number of samplings in each trial is 200.}
\label{XOR-sim}
\end{figure}
Fig.~(\ref{XOR-sim}) shows an example of learning XOR logic gate with physical neurons through the protocol described above.



\subsection{Continuous time learning}
\begin{figure}
    \centering
    \includegraphics[scale=0.7]{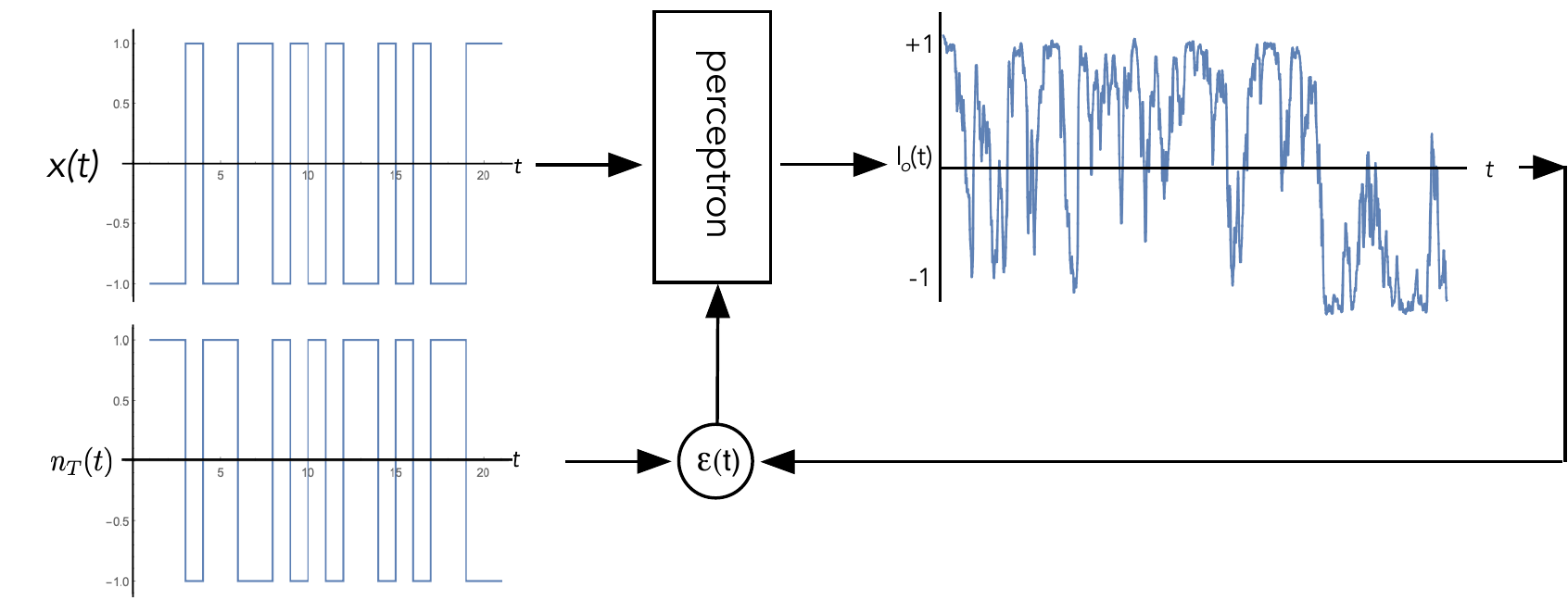}
    \caption{A feedback learning scheme for time-series data for the case of learning NOT.}
    \label{continuous-feedback}
\end{figure}

We now consider the  case in which the training data and labels are time-series data. Hence, the training data becomes a deterministic signal $x(t)$ as do the labels $n_T(t)$. The output of the perceptron is now a stochastic signal $n(t)$ as is the error function 
\begin{equation}
    \epsilon(t) =\frac{1}{4}(n(t)-n_T(t))^2.
\end{equation}
As the weights are chosen by a feedback process that depends on $\epsilon(t)$, the weights are likewise a stochastic function $\vec{w}(t)$. The objective now is to find stochastic differential equations that govern $n(t)$ and $\vec{w}(t)$. The overall scheme is depicted in Fig. (\ref{continuous-feedback}). 

In the absence of learning/feedback,  we assume that the initial weights satisfy an Ornstein-Uhlenbeck process with uniform decay plus white noise.   The stochastic differential equation is 
\begin{equation}
    d\vec{w}(t) = -\gamma_w\vec{w}(t) dt +\sqrt{D} \vec{dW}(t),
\end{equation}
where $\vec{dW}=(dW_1, dW_2, \ldots dW_N(t))$ are independent Wiener increments, $\gamma_w$ is a decay rate, and $D$ is the diffusion constant. The corresponding steady state distribution in weight space is the Gaussian
\begin{equation}
    P_{ss}(\vec{w}) =(2\pi D)^{-n/2} e^{-\gamma_w \vec{w}\cdot\vec{w}/2D}.
\end{equation}
This initialises the weights. Learning is now the process of engineering feedback to modify the stochastic dynamics on weight space so that the distribution becomes strongly peaked on the weights required to implement the function, in this case the NOT gate. This  is analogous to `cooling' in weight space. 

\begin{figure}
    \centering
    \includegraphics[scale=0.4]{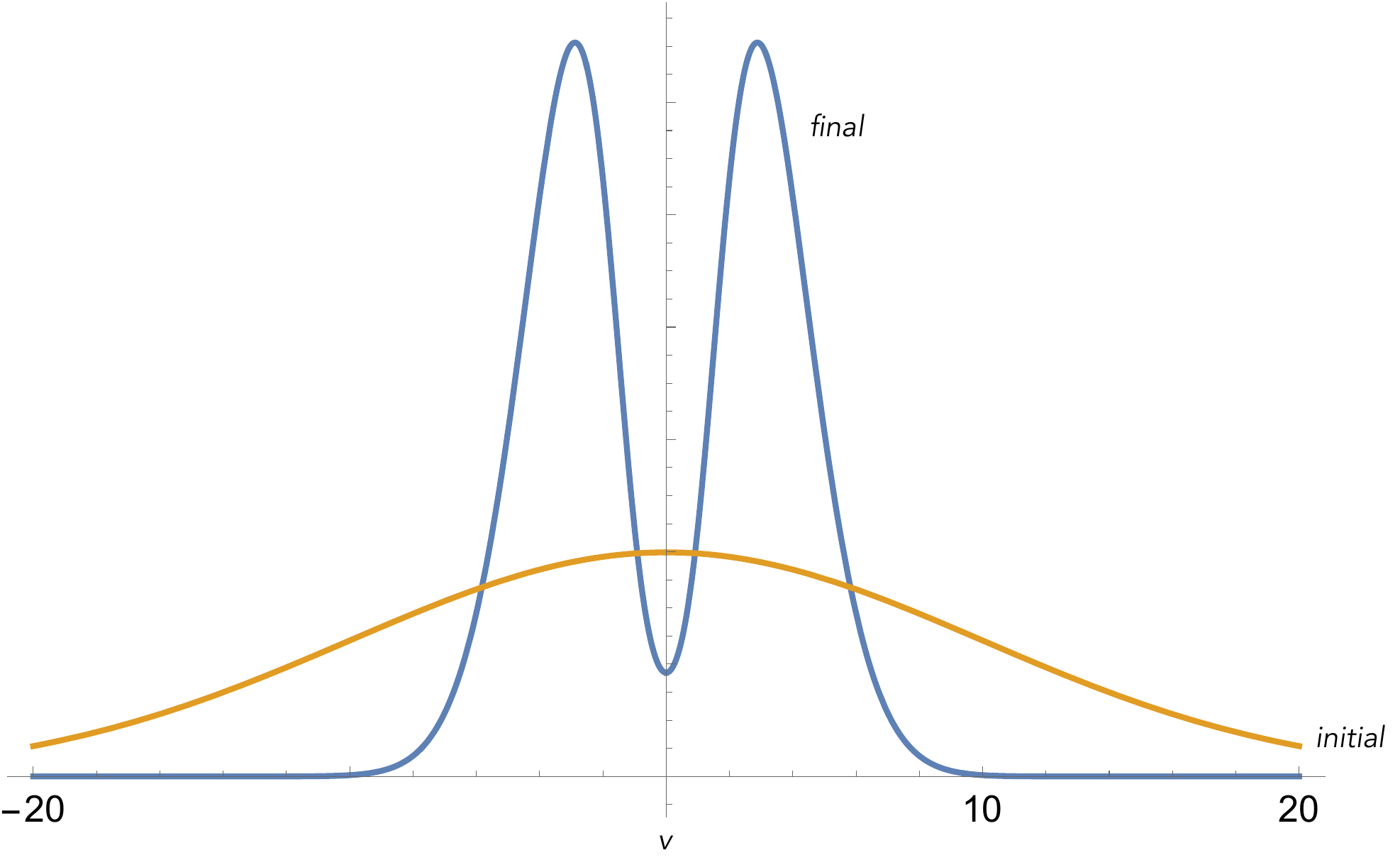}
    \caption{The initial and final (after learning) weight distribution. }
    \label{ss-dis-weight}
\end{figure}

As an example, we will consider the NOT gate once more. We will use the partially observed markov process discussed in section (\ref{partial-observe}) with a filtered observed current $I_o(t)$ and the state of the switch defined by the conditional mean $\bar{n}_c(t)$. 
The  underlying stochastic dynamical process is
\begin{equation}
    d\bar{n}_c(t)= \mu(1-\bar{n}_c(t))/2-\nu (1+\bar{n}_c(t))/2 -\sqrt{\Gamma/2}(1-\bar{n}_c(t)^2)dW,
\end{equation}
and the corresponding observed process is 
\begin{equation}
    dI_o(t) = -r^2 I_o(t)dt + r\kappa \bar{n}_c dt+r\kappa\sqrt{2/\Gamma}dW. 
\end{equation}
The transition rates are given by 
\begin{eqnarray}
   \mu(w,t) & = & \frac{\gamma}{1+e^{-\beta A(w,t)}},\\
   \nu (w,t) & = & \frac{\gamma e^{-\beta A(w,t)}}{1+e^{-\beta A(w,t)}},\\
\end{eqnarray}
where $A(w,t) = wx(t)+b$. The stochastic dynamics of the weights due to feedback can be deduced from the discrete case, Eq.(\ref{delta-weight}) 
\begin{equation}
    dw=L x(t)n_T(t)(1-\bar{n}_c(w,t)^2)dt-\gamma_w w  dt+\sqrt{D} dW,
\end{equation}
where $L=\eta\beta/4$ is the learning scale and we have included the intrinsic stochastic dynamics of the weights. The dependence of $n$ on the weights has been made explicit.  This equation tells us that the stochastic process for the weights has a very complicated drift term. As learning proceeds we expect $n(t)$ to approach $\pm 1$ and the increments in the weights becomes very small.   The increment is always negative for the NOT gate as $x(t)n_T(t)=-1$. In this case of NOT, we find that the stochastic differential equation for the weights becomes
\begin{equation}
\label{weight-sde}
    dw=-\frac{L}{\cosh(\beta w x/2)^2}dt-\gamma_w w  dt+\sqrt{D} dW.
\end{equation}
This is a Smoluchowski process~\cite{Gardiner} in the dissipative potential function
\begin{equation}
    V(w) =\frac{\gamma_w}{2} w^2 +\frac{2L}{wx}\tanh(\beta wx/2),
\end{equation}
where $x=\pm 1$ is the training signal. This means the second term is always positive. At the end of learning, the steady state distribution  for the weights is then  
\begin{equation}
    P_{ss}(w)={\cal N} e^{-\frac{V(w)}{2D}},
\end{equation}
where ${\cal N}$ is a normalisation constant.
The initial mean weight is zero and the variance is $\frac{D}{2\gamma_w}$. After learning, the stationary distribution is peaked near the fixed point of the deterministic part of Eq.(\ref{weight-sde}), thus the fixed points satisfy 
\begin{equation}
  {\rm sech}^2(v)+a v =0,
\end{equation}
where we have made the change of variable $v=\beta w/2 $ and 
$
    a=2\gamma_w/(L\beta^2)
$.
There are two cases to consider depending on the sign of $x$. Clearly as $a$ becomes large, the steady state weight approaches zero from either side. 
The variance of the final distribution can be determined by expanding the potential function to quadratic order around the fixed point.  In Fig.~(\ref{ss-dis-weight}), we plot the initial steady state distribution and the final steady state distribution of the weights. 
We see that the distribution is highly non-Gaussian and the variance is reduced substantially (one bit for the double peaks and a substantially reduced width for each peak) . This means that the effective temperature of the weights has been reduced by learning. Learning is analogous to cooling using feedback. It is interesting to ask how this view is manifest in a deep learning network trained by back propagation.  In back propagation, the feedback effectively cools the weights in each layer starting from the output and working back to the input. This suggests that a physical implementation of back propagation could be based on an analogy with heat propagation in an extended system, governed by a spatial diffusion equation.

\section{Thermodynamics of the perceptron}
\label{perceptron-thermo}

The weights control bias forces acting on the physical device, and  work is done by/on the activation switch as these forces change. We will define a single \emph{trial}, as one round of weight updates. This is not a single sampling of the switch. As we saw in the previous section, we need to sample the switching probability, with fixed data, over many time steps before updating the weights.    

The number of single runs required in each epoch depends on the value of $\beta A(\vec{w})$.  If we choose initial weights such that $\beta A(\vec{w})$ is far from zero, the probability of switching in each run is very small and we will need very many single runs to form a useful epoch for a single trial (see Fig.~(\ref{Bernoulli-example})).  A good initial  choice for the weights is thus a random distribution centred on the origin. This would make the learning rate very slow.  The width of this distribution is proportional to $1/\beta$. If $\beta>>1 $, only those weights near the origin will likely switch in a single sample. 

From trial to trial the work done/by the changing weights is a random variable as the time taken to switch is a random variable.  When a switch does occur, energy is dissipated as heat in a classical device or spontaneous emission in a quantum device, and so the energy dissipated per trial is also a random variable. We thus need to find a relationship between the fluctuating work done and energy  dissipated as a function of time. This is the subject of the field of statistical thermodynamics~\cite{Halpern_2015}. 

The change in the average energy of the switch is given by 
\begin{equation}
    \Delta \bar{E}= E_0(\Delta p_1(\vec{w})- \Delta p_0(\vec{w})).
\end{equation}
Using Eq. (\ref{two-state-ft}), we find
\begin{equation}
\label{energy-error}
     \Delta \bar{E}=-2\eta n_TE_0 \Delta \bar{\epsilon}. 
\end{equation}
This depends on $n_T$ and thus could be positive (work done on the perceptron)  or negative (work done by the perceptron). 
As the system learns, the change in the error per trial falls and the change in energy per trial also falls. In other words, when the change in error is small, the power required to learn is small. 

Let us now consider the change in Shannon entropy per step. This is given by
\begin{equation}
    \Delta S= \left [\Delta p_1(\vec{w})\right ]\ln \left (\frac{p_0(\vec{w})}{p_1(\vec{w})}\right ).
\end{equation}
Using Eq. (\ref{sigmoid}), we find that 
\begin{equation}
    \Delta S=\eta n_T\beta A(\vec{w})  \Delta \bar{\epsilon}.
\end{equation}
The change in entropy per step is also proportional to the change in error.  As learning proceeds, the entropy changes less.  Furthermore, the rate of change of entropy is proportional to $\beta$. 

In the thermally activated case, for which $\beta =(k_B T)^{-1}$, the change in the Helmholtz free energy ($\Delta F = \Delta E-k_B T\Delta S$) per trial is also proportional to the change in the average error. We conclude that when learning is successful, the change in the Helmholtz free energy per trial is minimum. In other words, when learning is complete the average work done on/by the perceptron is a minimum.  In the quantum case $\beta$ is not an inverse temperature but rather determined by quantum parameters such as a tunnelling rate.

Goldt and Seifert~\cite{GS} have found an equivalent thermodynamic constraint on information theoretic measures of learning efficiency for classical learning machines. This could easily be generalised to the quantum case. 

The relationship between rate of change of energy and rate of change of error in Eq. (\ref{energy-error}) is similar in spirit to Landauer's erasure cost~\cite{landauer1961irreversibility} in that it relates a physical quantity to an information theoretical quantity. It holds in both the classical and quantum regime.

\section{Spiking Neural Networks}

The classical perceptron originated in early dynamical models for neurons in the mammalian brain. However, unlike the preceptron model, neurons exhibit self-sustained oscillations. Deterministic and stochastic models are found in many textbooks and monographs~\cite{DayanAbbott, LaingLord,HH}. Underlying these models is the concept of a Hopf bifurcation whereby a dissipative fixed point becomes unstable due to a parameter variation leading to a self-sustained oscillation called a limit cycle~\cite{GuckenheimerHolmes}. In this section, we will discuss how Hopf bifurcations can be used to build an activation switch with better noise characteristics than the fixed point models like the noisy double well discussed above. 

Our approach is not the traditional approach to spiking neural networks (SNNs), however, we believe it is better physically motivated while exhibiting many of the key features found in neuroscience models.  It must also  be acknowledged that the algorithmic implementation of SNNs for learning is not as developed as that for artificial neural networks used in deep learning. A good introduction to SNN algorithms may be found in~\cite{SNN}. 

The over-damped double well potential enables an activation switch based on the zero dimensional attractors of the steady state, and noise pays a crucial role in the  stochastic switching required for learning.  It might be desirable to tune this noise for optimum performance.  In the classical case, the noise required for switching depends on the temperature, and in the quantum case it depends on a quantum tunnelling. This makes it difficult to tune the noise {\em in situ}.  However, if we move from zero dimensional attractors to one dimensional attractors --- limit cycles --- we can tune the noise by changing the driving power in both the classical and the quantum case. We now explain how this works.

We are all familiar with limit cycles as they are how clocks work~\cite{Milburn-clocks}. All clocks are dissipative devices; energy lost to the environment will eventually cause them to slow down until they exhibit fluctuations around an equilibrium position.  For this reason energy must be constantly supplied by a steady power source.  
Good clocks are necessarily dissipative. This is also true for quantum clocks~\cite{Erker}. 

A quartz clock based on a Schmidtt trigger~\cite{Milburn-clocks} provides a good example to introduce neuronal spiking. The key feature of these models (such as Hodgkin-Huxley~\cite{HH}) is a nonlinear coupling between transition rates in an in-homogeneous Markov process (see Eq. (\ref{classical-me})) and another dynamical variable.  The quartz clock (and the pendulum clock, for that matter) is described by the equations
\begin{eqnarray}
\label{quartz-lc}
\dot{v} & = & \mu(x)(1-v)- \nu(x)(1+v), \\
\dot{x} & = & y,\\
\dot{y} & = & - x -\kappa y+\chi (v-\epsilon),
\end{eqnarray}
where
\begin{eqnarray}
\mu(x)= \gamma e^{- x},\\
\nu(x)= \gamma e^{ x},
\end{eqnarray}
and $\epsilon$ is a bias voltage. The units have been chosen such that the period of the free oscillator is unity.  This variable is coupled to a damped simple harmonic oscillator with natural frequency $\omega$ and dissipation rate $\kappa$. The variable $x$ controls the bias conditions for the Schmidtt trigger.  This model also describes the semi-classical limit of a nano-mechanical oscillator coupled to a tunnelling quantum dot~\cite{Milburn-clocks}.  Clearly this is not a good biophysical model for a physiological neuron, but it will suffice to illustrate an important general principle: fluctuations in the time between spikes.  This is a key feature of spiking physical neural networks and is essential for learning. 

\begin{figure}
\centering\includegraphics[scale=0.5]{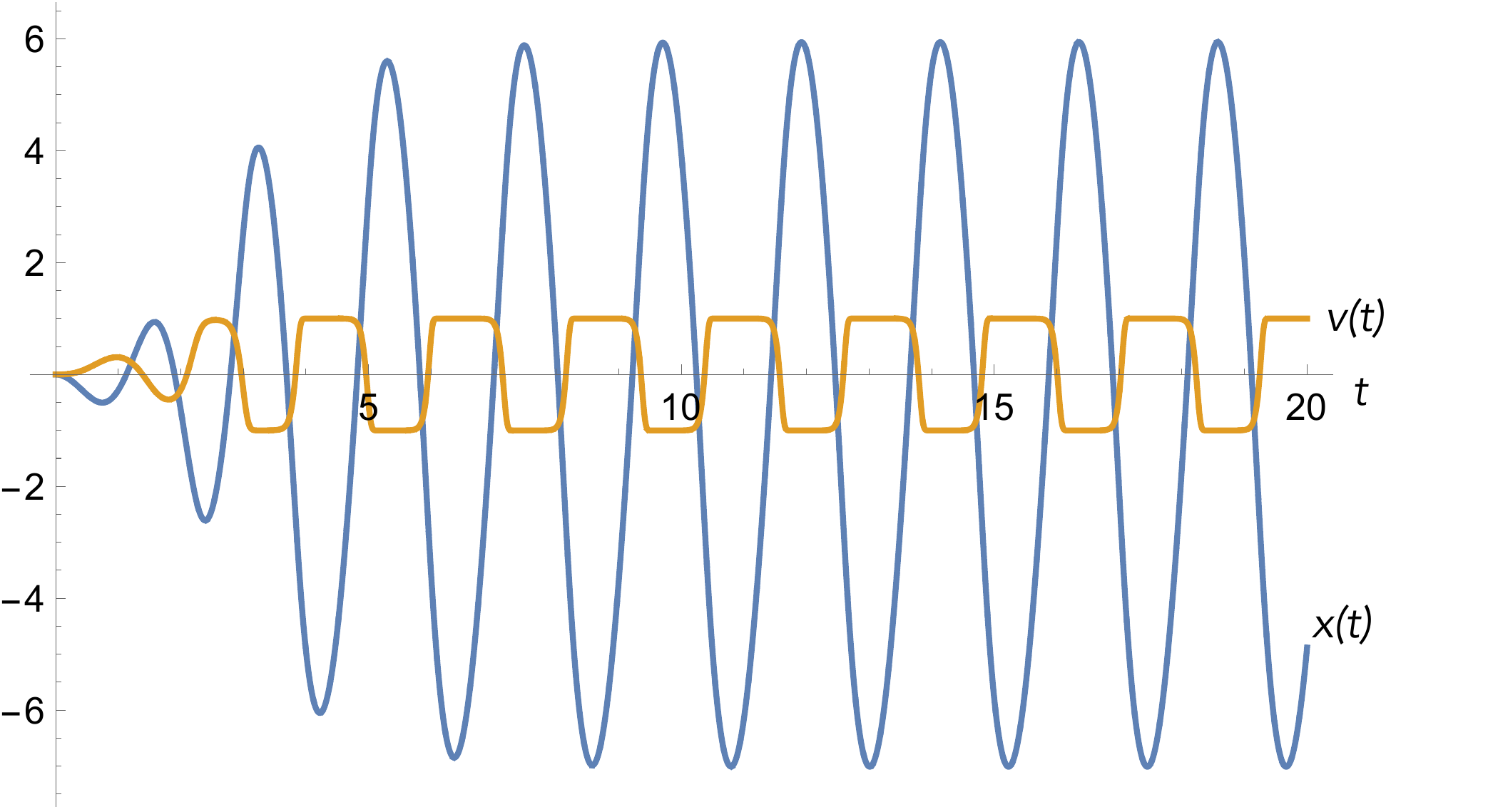}
\caption{The dynamics for  $v(t)$. and $x(t)$ when the system parameters are chosen to give a limit cycle for the deterministic dynamics.   $\gamma=1.0,  \kappa =1.0$,  $\chi=40.0$, $\epsilon=0.1$ }
\label{quartz}
\end{figure}

When $\epsilon=0$, the origin in the dynamical variables is a stable fixed point. As $\epsilon$ is increased, the origin goes unstable and a limit cycle forms for $\chi$ above a critical value $\chi_h$ (see~\cite{Wahyu}). For large values of $\epsilon$, the limit cycle disappears and a stable fixed point returns.  In Fig.~(\ref{quartz}), we plot the dynamics for the variables $x(t), v(t)$ on the limit cycle.  
The conditions for a fixed point or limit cycle can be controlled by changing the value of $\chi$ or $\epsilon$. For example, if we let the coupling constant $\chi$ be modulated by the integrated signal from the output of another spiking neuron, eventually, it will reach the critical value and the target spiking neuron will begin to oscillate in response.  If the integrated signal falls below the critical value, the target spiking neuron will cease to oscillate.  This provides a way to train a spiking perceptron based on this model. 

It is evident that for these parameters, the harmonic oscillator is subjected to a periodic kick of magnitude $\pm\chi$, phase shifted with respect to the sustained oscillations in $x$. This is similar to a dynamical description of a pendulum clock with an anchor and wheel escapement mechanism \cite{Milburn-clocks} where the variable $v$ describes the highly dissipative dynamics of the escapement. 

We find the limit cycle by defining $r,\psi$, by $x=r\cos\psi$ and $y= r\sin\psi$. On the limit cycle, we write $v(t)=\tanh(r\sin(\psi-\psi_0) )=-\tanh[\sin\psi_0 x-\cos\psi_0 y]$ where the phase shift $\psi_0$ is determined below.  If the limit cycle is large, $r=r_*$, we can approximate $v(t)=-{\rm sign}[\sin(\psi-\psi_0)]$.
 The time scale of the decay onto the limit cycle (the $r(t)$ dynamics) is slower than the oscillation frequency on the limit cycle (the $\psi(t)$ dynamics), and we can average over the oscillatory fast time scale which is equivalent to averaging over $\psi$ from $-\pi-\psi_0$ to $\pi-\psi_0$.  The equation for the radial coordinate after averaging is
\begin{equation}
\label{radial-motion}
\dot{r} = -\frac{\kappa r}{2}+\frac{2\chi}{\pi}\cos\psi_0.
\end{equation}
There is a stable fixed point in the radial coordinate given by
\begin{equation}
\label{radial-fp}
r_*=\frac{4\chi}{\pi\kappa}\cos\psi_0.
\end{equation}
The dynamics for the phase is 
\begin{equation}
\label{phase-dynamics}
\dot{\psi} = -\omega+\frac{2\chi}{r\pi}\sin\psi_0.
\end{equation}
If the limit cycle is large so that $\dot{\psi}=-1$,
 we find that 
\begin{equation}
    \cos\psi_0 \approx \frac{1}{r_*}\ln(r_*/\gamma).
\end{equation}
In the limit $\gamma >>1$, we see that $\psi_0\rightarrow \pi$. This is not surprising. Inspection of the equation of motion for $v$  indicates that when $\gamma$ is large we expect the variable $v$ to switch almost instantaneously when $x$ changes sign.

\begin{figure}
    \centering
    \includegraphics[scale=0.5]{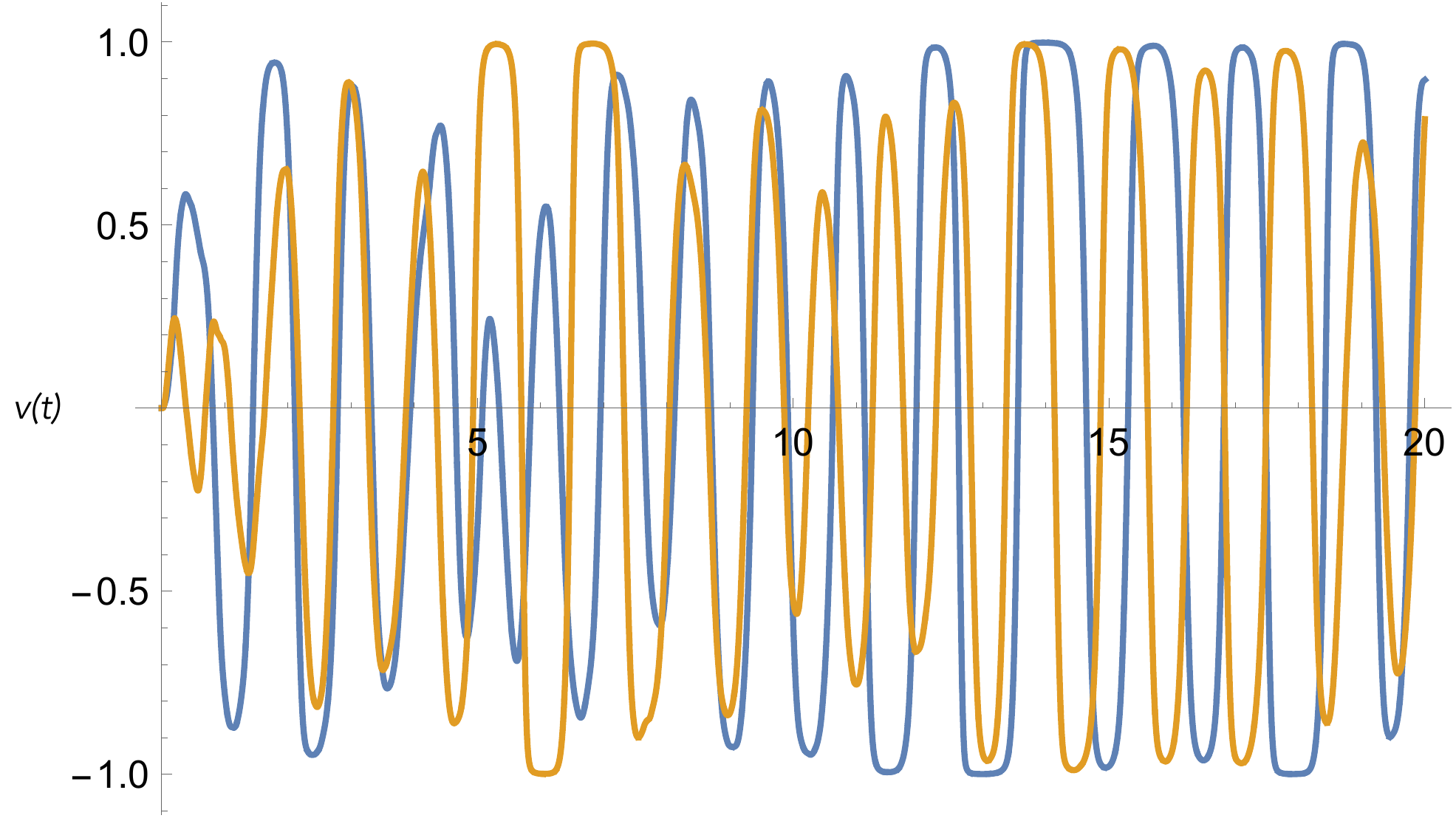}
    \caption{Two sample trajectories for $v(t)$ with thermal noise included.  Fluctuations in the period resulting from phase diffusion on the limit cycle are evident. Parameters are: $\gamma=10, \chi =40, \kappa=1.0,\sqrt{\sigma}=5.0$. }
    \label{obs-quartz}
\end{figure}

We now need to consider the effects of noise. As the oscillator is damped, the fluctuation dissipation theorem indicates that there must be a thermal Langevin force in the momentum variable $y$. The  Ito stochastic differential equations are now 
\begin{eqnarray}
\label{noisy-quartz-lc}
dv & = & \mu(x)(1-v)- \nu(x)(1+v), \\
dx & = &  y dt,\\
dy & = & - x dt -\kappa y dt+\chi (v-\epsilon)+\sqrt{\sigma} dW dt,
\end{eqnarray}
where $\sigma$ is proportional to temperature. 
In Fig(\ref{obs-quartz}), we plot two sample trajectories for $v(t)$.

In the limit of large $\eta$, the dynamics is given by  Ito stochastic differential equations. A detailed discussion is  given in~\cite{Milburn-clocks}. On the limit cycle this results in phase diffusion described by 
\begin{equation}
d\psi = - dt+\frac{2\chi}{r_*\pi}\sin\psi_0 dt +\frac{\sqrt{\sigma}}{r_*} dW(t).
\end{equation}
The significant feature here is that the rate of phase diffusion {\em decreases } as the size of the limit cycle increases. This means that we can control the noise, without changing the temperature, simply by increasing the coupling strength, $\chi$. This offers a useful design tool in building artificial spiking perceptrons.

\begin{figure}
\centering
\includegraphics[scale=0.5]{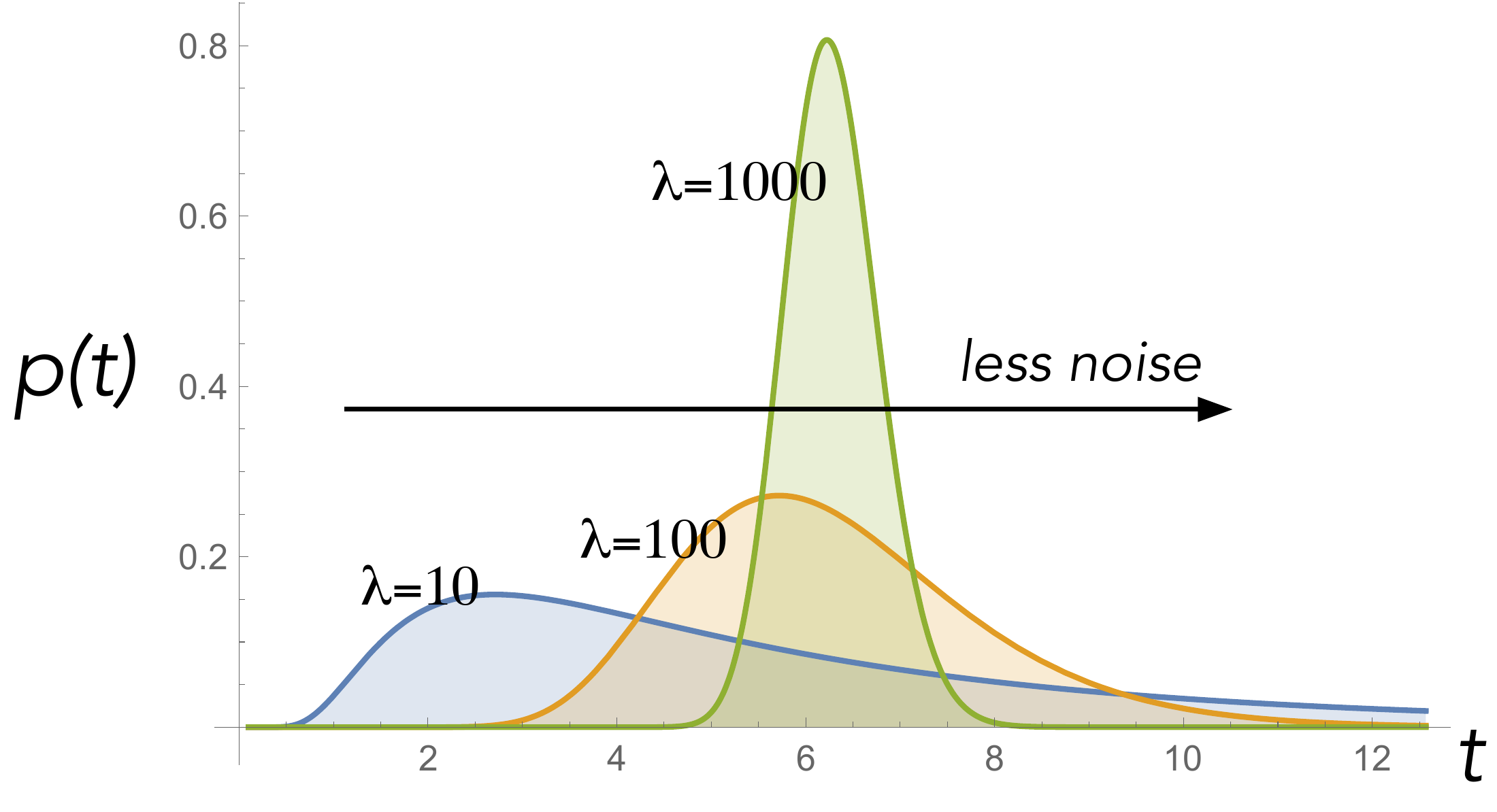}
\caption{The inverse Gaussian, or Wald, distribution for the periods on a noisy limit cycle. The averaging period is $2\pi$ and three spread parameters are illustrated to control the amount of phase diffusion. }
\label{wald-dist}
\end{figure}

The phase noise leads to fluctuations in the cock period. This is a `first passage time' problem: the time taken for the phase to advance by $2\pi$ is a random variable. This time is the fluctuating period of the clock. If the phase obeys the Ito stochastic differential equation, the probability distribution is given by the inverse , or Wald, distribution~\cite{Aminzare} 
\begin{equation}
    P(T)= \sqrt{\frac{2\pi\mu}{\sigma T^3}}\exp\left [-\frac{r^*(2\pi- T)^2}{2\sigma T}\right ].
    \end{equation}
    The mean and variance are
\begin{eqnarray}
   {\cal E}[T] & = & 2\pi,\\
   {\cal V}[T] & = & \frac{2\pi\sigma}{r^*}.
\end{eqnarray}
The inverse Gaussian is usually written in terms of the mean and the spread defined by $
  \lambda =\frac{ {\cal E}[T]^3}{{\cal V}[T]},
$
with units of seconds. The distribution is illustrated in Fig.~(\ref{wald-dist}) for increasing values of the spread parameter for $\omega=1$. 

 The energy dissipated by the oscillator component of the system of equations is proportional to the energy in the case of viscous damping
$
     \dot{E}=-\kappa E.
$ 
 As the energy function in dimensionless units is $E=(x^2+y^2)/2 $, we see that the energy is proportional to $r^2$. The rate of energy dissipated is thus proportional to the square of the size of the limit cycle, $r_*$.  This means that when the phase diffusion is small, the rate of energy dissipation is large. The thermodynamics of limit cycles implies that fluctuations in the period can be kept small at the expense of power dissipated.

\subsection{Training a spiking perceptron}
We will assume that the output of the spiking perceptron is simply the variable $v(t)$ as this is always of fixed amplitude. More generally, we will take the actual output to be the current resulting from a continuous weak observation of $v(t)$. For example, in a quantum nanomechanical realisation~\cite{Wahyu} the current would correspond to the current through an SET monitoring the occupation of the dot.

We modify the observed process slightly to count cycles, so we  shift the variable $v$ in the current equation so that the observed process is between $0$ and $1$  
\begin{equation}
dJ=-k^2 J dt+kw(1+v)dt+k\sqrt{\sigma}dW.
\end{equation}
The observed process defined here is an effective `leaky integrate and fire' model. The value of $k$ determines the leakage rate: A small value indicates almost instantaneous response to changes in $v$. The signal-to-noise ratio is determined by $\sigma$.   The variable $w$ is the weight. As a final embellishment, we include the possibility that the coupling to the input signal turns off after some time, so we write
\begin{equation}
\label{LIF}
dJ=-k^2 J dt+kw h_\tau(1+v)dt+k\sqrt{\sigma}dW,
\end{equation}
where 
\begin{equation}
    h_\tau(t) =(1+e^{\alpha(t-\tau)})^{-1},
\end{equation}
in which $\alpha>0$ determines how fast the turn-off occurs. We regard Eq. (\ref{LIF}) as describing the internal dynamics of each LIF unit in a neuron: in effect it describes a synapse.  

The question of how best to train an SNN model is unsettled. There is as yet no standard model like the back propagation algorithm used in deep learning networks. SNN models with application to machine learning are reviewed in \cite{Yamazaki}. Many of these models are based on the general idea of the classic `leaky integrate and fire model'.   
Our treatment derives from this but we incorporate the insight, discussed in the previous section, that spiking neurons are essentially clocks. The idea is to monitor the total integrated current (Eq.(\ref{LIF})) over some fixed time $\tau$ of a post-synaptic neuron.  When this surpasses some threshold, a nonlinear switch is designed to trigger a Hopf bifurcation in the post-synaptic neuron when the threshold is reached.  The effect of noise enters due to phase diffusion on the limit cycle which leads to fluctuating periods from one cycle to the next. This implies that the input power to the switch is a random variable as it is for the physical perceptron model previously discussed.

\begin{figure}
    \centering
    \includegraphics[scale=0.5]{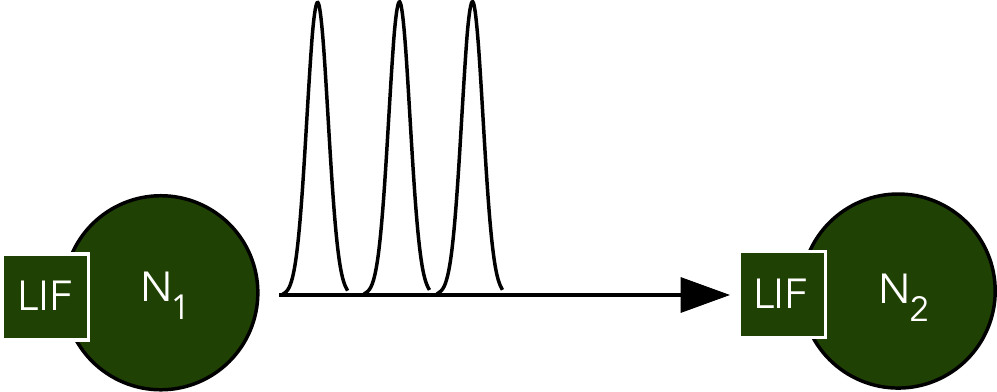}
    \caption{The leaky integrate and fire model of a spiking neural network of two neurons. The first neuron, the pre-synaptic neuron,  has undergone a Hopf bifurcation and produces an oscillatory output. The number of spikes in some time $\tau$, is set by design in the  leaky integrate-and-fire (LIF) switch. The LIF component of the second neuron, the post-synaptic neouron,  counts the number of oscillations and if a threshold is reached, it undergoes a Hopf bifurcation and produces an oscillatory output. } 
    \label{SNN-scheme}
\end{figure}

We will assume that the integration time is set by the receiving neuron, see Fig.~(\ref{SNN-scheme}).
That is to say, the leaky integrate-and-fire dynamics, (Eq.(\ref{LIF})), is part of the input to each neuron.  The integration time, $\tau$, is fixed for every neuron.  The number of cycles recorded is the random variable $N(\tau)$. We then set a threshold, $N_{th}$, such that the probability to fire is 
\begin{equation}
    p(\vec{w})=(1+e^{-\beta (N(\tau)-N_{th})})^{-1},
\end{equation}
where $\beta$ is a positive real parameter. If the switch fires, the next neuron is biased so as to undergo a Hopf bifurcation and start spiking.  

\begin{figure}
    \centering
    \includegraphics[scale=0.5]{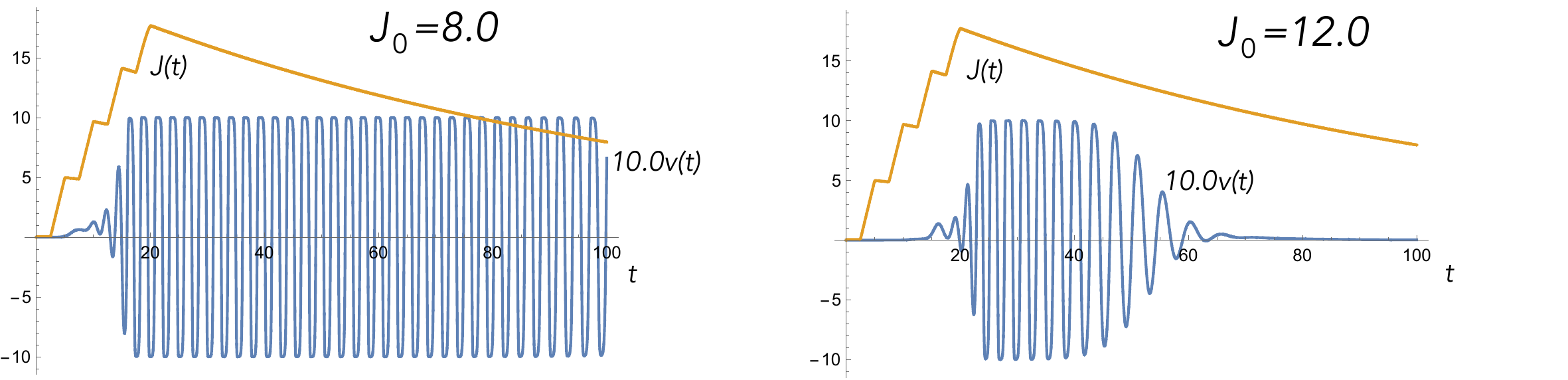}
    \caption{A plot of the integrated current $J(t)$ and the output $10*v[t]$ for a post-synaptic neuron driven by a pre-synaptic neuron with a period of $T=5.0$ with two values for the threshold current $J_0$. Other parameters are $\chi = 20, \kappa=1,\gamma=1.0, L=0.01,\beta=1.0, \lambda=10.0,\tau=20.0, w=1.0 $}
    \label{feed-forward}
\end{figure}

In terms of the spiking neuron model we have used here, we can implement a deterministic version of this protocol by modulating $\chi$
\begin{equation}
    \chi(t)=\chi\left (1+e^{-\beta(J(t)-J_0)}\right )^{-1},
\end{equation}
where $J_0$ is the threshold value. 
In Fig. (\ref{feed-forward}), we illustrate this switching using a pre-synaptic signal of the form $m(t)=\tanh[\lambda \sin(2\pi t/T)]$ where the period of the input signal is $T$ and we take two values of the threshold.  

\begin{figure}
    \centering
    \includegraphics[scale=0.4]{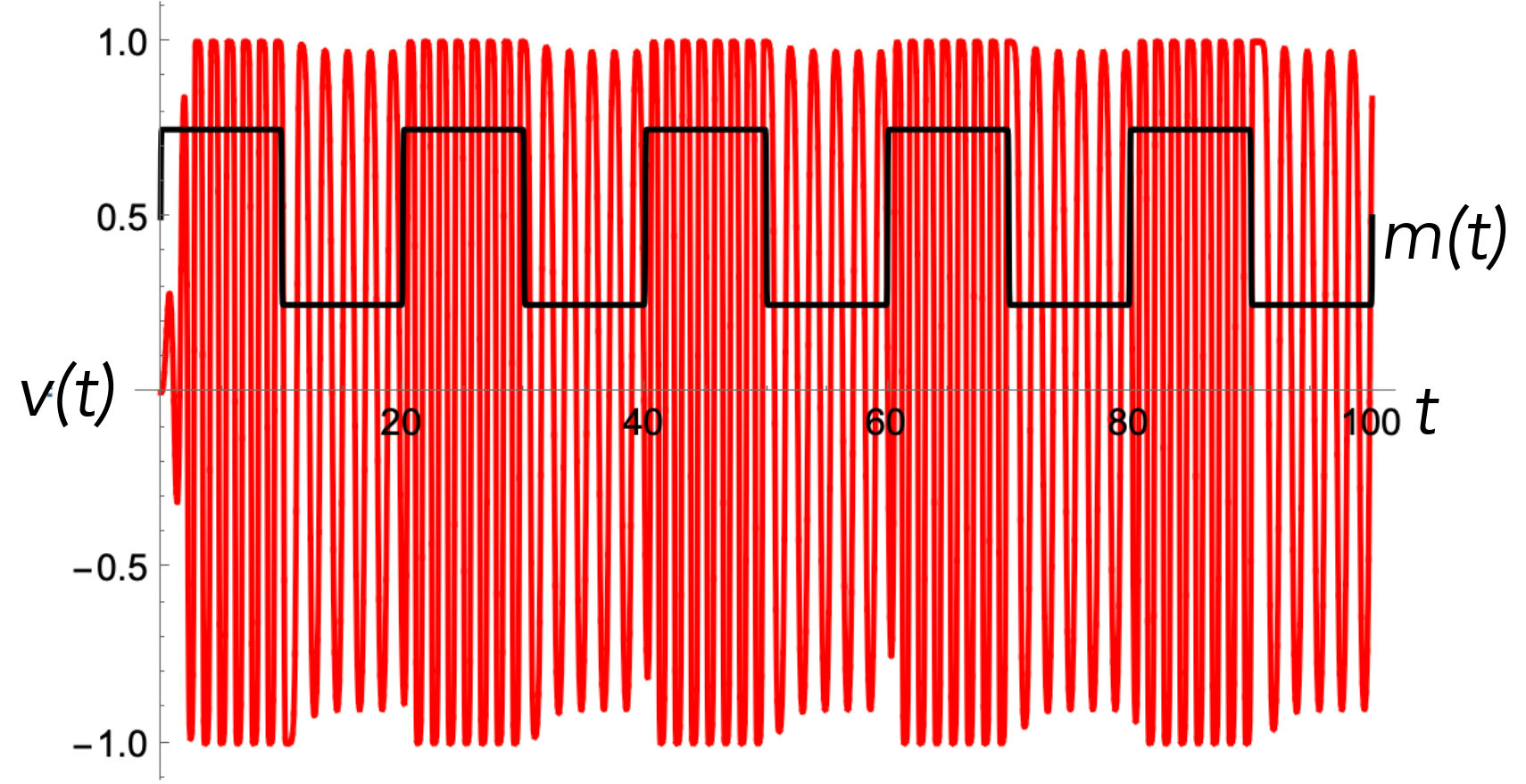}
    \caption{A spike rate temporal code for a binary signal. The output $v(t)$ when the coupling constant is modulated.}
    \label{spike-rate-code}
\end{figure}

How do we encode the data to train spiking neurons? The simplest is rate coding, equivalently, period encoding. As an example, consider the case of binary data. We chose two clocks each with a different mean period. In the above example, the period depends on the coupling constant $\chi$ (see Eq. (\ref{phase-dynamics})).  This is equivalent to frequency modulation. As an example, we let the coupling constant become time dependent 
\begin{equation}
 \chi(t) =\chi (1+\theta M(t) ), 
\end{equation}
where $M(t)$ is a square wave signal between $\pm 1$ encoding a time-series binary signal, and $0<\theta \leq 1$.  In Fig.~(\ref{spike-rate-code}), we give an example by solving the dynamical equations (without noise) for $v(t)$ with  $\chi(t)=(\chi/2)(1+0.5. {\rm sign}\left [\tanh[\sin(2\pi t/20)]\right ]$.

In a spiking perceptron network, weights are implemented in the LIF component of each neuron by summing over all incoming signals
\begin{equation}
dJ_k=-k^2J_k dt+ k\sum_l w_{kl}h_{\tau,l} (1+v_l)dt+k\sqrt{\sigma}dW,
\end{equation}
where the weights are $w_{kl}$. An alternative to the rate code is spike train delay plasticity (STDP)~\cite{Yamazaki}. In our model, this corresponds to feeding back onto the switch-off function $h_\tau$ to change the integration time $\tau$.

\subsection{Quantum oscillatory neural networks}
 There are many examples of limit cycles driven by quantum noise at very low temperature. The standard example is a laser. This is a device that is pumped by an external supply of energy, with a nonlinear amplification process and dissipation. For example, in a gas of atoms  the pump could be a source of thermal radiation  that pumps atoms from a ground states to an excited state so as to generate a population inversion between two dipole allowed transitions.   The amplification process is based on stimulated  emission on that transition and the dissipation is loss of stimulated photons through an optical cavity resonant with the stimulated radiation field. There is a critical rate at which energy is pumped into the system below which the radiation emitted from the cavity is predominantly due to spontaneous emission with a mean field of zero and thermal photon statistics. Above that critical pumping rate, the system undergoes a transition to emit a coherent radiation field with a non-zero mean field and Poissonian photon statistics~\cite{WallsMilburn}. 
 
 The quantum nature of the system is reflected in the appearance of spontaneous emission. A fully quantum description is well known and shows how spontaneous emission noise leads to phase diffusion of the field amplitude above threshold. The quantum theory shows that the phase diffusion rate decreases as the dissipation of energy increases.   There is a semi-classical description which reproduces the mean field response based on the van der Pol oscillator and so it is easy to see how this can be used to make a spiking perceptron. However, as the system operates at optical frequencies it is difficult to count the number of limit cycles. The same problem is faced when using a laser limit cycle to make a clock. The solution to this is also well known and uses mode locked lasers~\cite{Milburn-clocks} and the same approach can be used to engineer spiking perceptrons. Another example of a quantum spiking perceptron can be based on the single electron shuttle~\cite{Wahyu, Waechtler}.  This has the advantage of being a nanoscale device consuming very low power and would thus be a promising technology on which to build efficient learning machines.  
 
 \section{Quantum kernel embedding}
 We have emphasised that all learning machines are dissipative and have indicated that this does not preclude quantum enhancements: dissipative quantum tunnelling allows activation switches to operate at very low temperatures and low power. One might get the impression that there is no role for non-dissipative unitary quantum devices in machine learning. This is certainly not the case. 
 
 The recent work of Huang et al.~\cite{Huang} highlights the role of quantum coherence in enabling a speedup via data representation using quantum states.   They give an example with an exponential quantum advantage. A key role is played by a quantum memory. We will describe another example that uses a very simple photonic quantum memory to evaluate a kernel. 
 
Kernel methods are a machine learning tool to evaluate inner products in high, or even infinite, dimensional feature space.
It has been suggested that quantum mechanics could be used to make this more efficient given that the probability distributions in quantum mechanics are given by inner products~\cite{PhysRevResearch.1.033159}. In a Raman quantum memory for single photon states inner products determine the probability of storing and retrieving the state~\cite{Kalb}. 

Single photon states are highly quantum states of light in which a single excitation of the field is superposed over many frequencies to create a pulse of light with one and only one photon~\cite{RSMilburn}. The average field of such a state is zero but the intensity is not. Single photon states can be created in a Raman single photon source and stored in a Raman quantum memory~\cite{KMS,RayWal}.  In such a source the temporal shape of the photon is determined by a classical pulse and this pulse may be carefully crafted using standard optical modulation methods. 

Single photon states are highly non-classical field states. They are defined by 
\begin{equation}
\label{single-photon-state}
 |\nu(t)\rangle =\int_{-\infty}^\infty \ d\omega \tilde{\nu}(\omega) a^\dagger(\omega)|0\rangle,
\end{equation}
where normalisation requires that
\begin{equation}
    \int_{-\infty}^\infty \ d\omega |\tilde{\nu}(\omega)|^2 =1,
\end{equation}
and the external field amplitude operator is given by 
\begin{equation}
    a(x,t)= \int_{-\infty}^\infty\ d\omega  e^{-i\omega (t-x/c)}  a(\omega).
\end{equation}
This is valid when the carrier frequency of all optical states is much greater than the bandwidth of those states. 
The state defined in Eq. (\ref{single-photon-state}) is a coherent superposition of a single excitation over many frequency modes. The physical meaning of the amplitude function $\tilde{\nu}(\omega)$ is seen when we ask for the detection probability per unit time of this photon for a point detector located at space time point $(x,t)$. This is given by 
\begin{equation}
    r(t)=\eta \langle  a^\dagger(x,t)a(xt)\rangle =\eta |\nu(t-x/c)|^2,
\end{equation}
and $0<\eta< 1$ is a detector dependent parameter called the efficiency and $\nu(t)$ is the Fourier transform of $\tilde{\nu}(\omega)$. In this case the resulting optical state is transform-limited. Thus transform-limited states are pure states; they have no more phase noise than is required by the uncertainty principle. 

A data element is a unit vector $\vec{\xi}$ with components, $\xi_j\in \{\pm 1\}$. 
Let $\{u_j(t)\}$, with $j=1,2,\ldots, N$, be a set of orthonormal temporal mode functions.  To encode this in a single photon state we define the temporal mode function $\nu(t)$
\begin{equation}
\label{input}
    \nu(t) =\sum_{j=1}^N \xi_j u_j(t).
\end{equation}
One easily checks that
\begin{equation}
    \int_{-\infty}^\infty\ dt |\nu(t)|^2 =1,
\end{equation}
so the state is normalised. We refer to $\nu(t)$ as the `write pulse` as it writes the data into the single photon state. This state can be created by a suitable control pulse in a Raman source. 

A single photon detector can be realised using a single photon Raman memory followed by atomic state detection~\cite{james_atomic-vapor-based_2002}. The probability amplitude to store the photon is determined by a classical control pulse  we call the write pulse. This is conveniently described by a projection operator onto a controllable single photon pulse
\begin{equation}
    \Pi_{\alpha(t)} = |\alpha(t)\rangle\langle \alpha(t)|.
\end{equation}
We now suppose the write field is given in terms of a unit weight vector, $\vec{w}$, with components $w_j$, as
\begin{equation}
    \alpha(t) =\sum_{j=1}^N w_j u_j(t),
\end{equation}
where we use the same set of orthonormal temporal mode functions.  
The detection probability for an arbitrary single photon state, $|\nu(t)\rangle$, is a function of  
\begin{equation}
    A(\vec{w})= \eta|\langle \alpha(t)|\nu(t)\rangle|^2,
\end{equation}
where $0< \eta\leq 1 $ is a (device dependent) dimensionless parameter called the quantum efficiency. The function $\alpha(t)$ is the `read pulse` as it is used to readout the single photon.  Then
\begin{equation}
A(\vec{w})  =\left |\int_0^\infty dt\  \alpha^*(t)\nu(t)\right |^2=|\sum_{k=1}^N w_k \xi_k|^2.
\end{equation}
This is zero  if $\vec{w}$ is orthogonal to $\vec{\xi}$, and unity if they are parallel. In this form, we see that the data encoding realises a single photon quantum kernel embedding.  The probability distribution for a detection event, $p_1$, is given by a nonlinear function (typically exponential) of $A(\vec{w})$, 
\begin{equation}
    p_1(\vec{w})=f( A(\vec{w})).  
\end{equation}
The inner product is evaluated by sampling the detection events. The ability to write the data into the pure state of a single photon and store the single photon in a quantum memory is what enables the inner product to be evaluated by sampling the detection probability. This gives a very energy efficient way of evaluating the required inner product without lots of matrix-vector multiplications.

\section{Physical learning machines and the free energy principle}
We have stressed the central role of thermodynamical concepts, particularly energy dissipation and free energy,  in the physical understanding of learning machines. In classical machine learning a number of   statistical  concepts have been introduced with similar mathematical constructions to thermodynamics but lacking a clear  connection to an underlying physical system. Indeed, the Wikipedia entry for the free energy principle begins with a heading \emph{Not to be confused with Thermodynamic free energy.}   It is our intention here to show a link between physical thermodynamics of learning machines and a statistical method used in reinforcement learning known as the free energy principle (FEP), primarily due to Friston~\cite{Friston}.  
We will show how the stochastic thermodynamic fluctuation theorems can connect physics with statistical methods. More work needs to be done to demonstrate this link in other contexts, such as Hinton's Helmholtz machine~\cite{Hinton}, and Alemi's and Fischer's work on therML~\cite{therML}. 

Friston's free energy principle (FEP) attempts to capture key statistical features exhibited by all learning agents, especially biological ones,  that must control features in their environment in order to maintain a metastable state far from thermal equilibrium. As originally stated the principle is a little ambiguous and has attracted diverse interpretations. We cannot review all these here, but instead focus on how one formulation of it can be related to the physics of learning machines. 

\begin{figure}
    \centering
\includegraphics[scale=0.5]{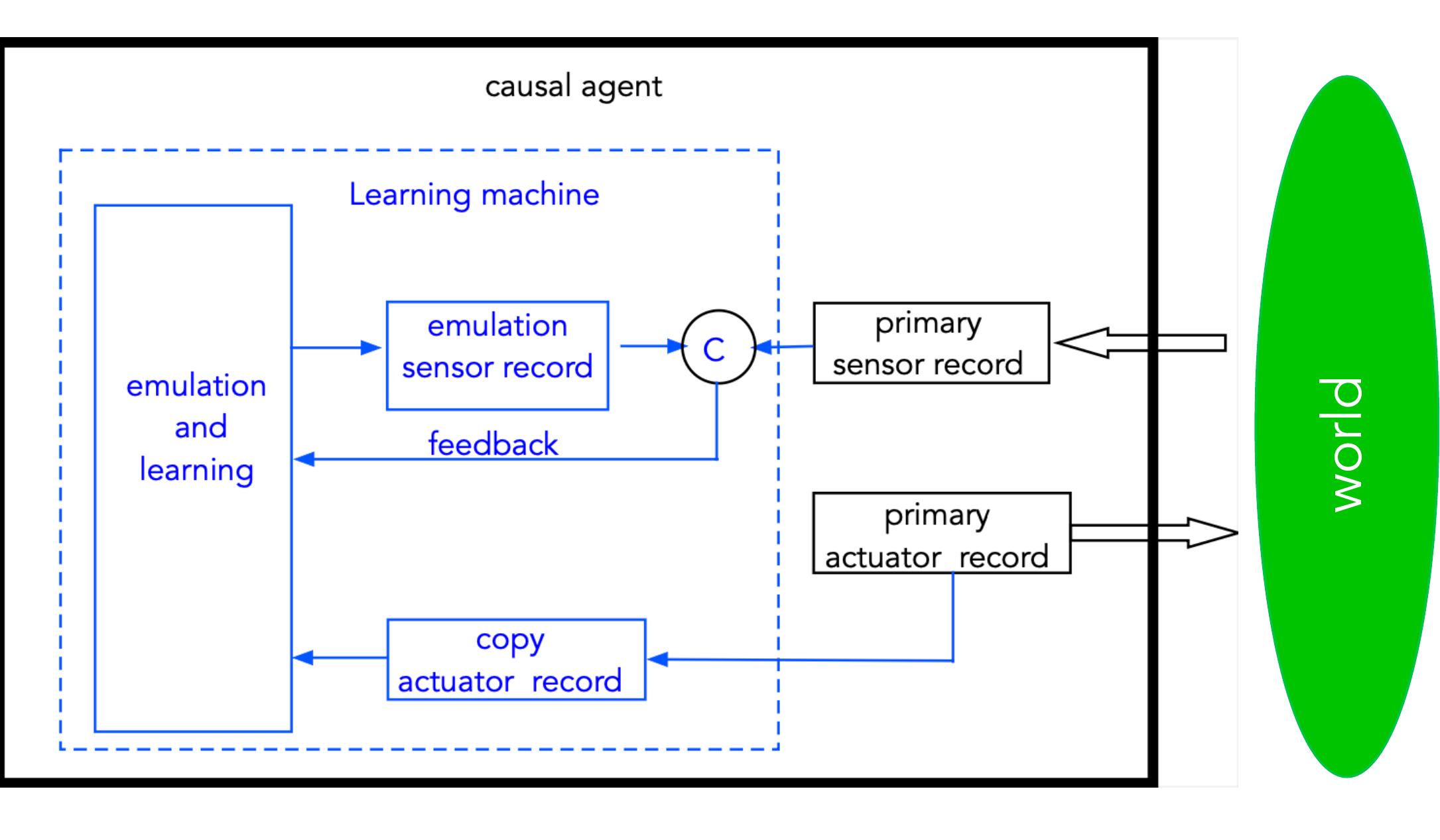}
    \caption{A schematic of a learning process based on a physical emulator with feedback. The agent acts on the external world using actuators and the world responds by acting on the agents sensors. }
    \label{learning-emulator}
\end{figure}

One way to formulate the FEP is in terms of an agent using actuators to change its environment, sensors to record how the agent responds to changes in the environment, and a learning emulator that seeks to make predictions of sensor records on the basis of the actions taken.  We call this emulation prediction learning (EPL).  The key feature of the emulator is that it is controlled by an internal feedback based on comparing actual sensor records and predicted sensor records to make the probability of disagreement (error probability)  as low as possible, see Fig.~(\ref{learning-emulator}).

Elsewhere, one of us has discussed the physical grounds of such a model~\cite{MilburnShrapnel2018}. A key feature of EPL is that it takes an  agent-centred perspective: the only states the agent `knows' are its internal sensor and actuator records as well as physical variables describing its learning emulator.  This makes it easier to discuss the case in which the world is fully quantum and the agent makes measurements upon it to update its sensor records. The agent does not need to refer to external states of the world itself. In this perspective the world is an `oracle'.    

We will assume the ELM is based on a physical perceptron network.  Learning then requires  updating of weights, by feedback. The weights encode a function $s=A_w(a)$ that maps actions to predicted sensations that, with low probability of error, match actual sensations.  In other words, once learning is complete the agent is rarely surprised by badly wrong predictions.    

If, after learning, the emulator is identical to the world system, the error will be very small.  This is a result that sometimes goes by the name of the `good regulator theorem'~\cite{good-regulator}. In general, this is unlikely to be the case. The agent only needs to learn  the function $s=A_w(a)$ with low probability of error and this could be achieved by physical perceptron networks that look nothing like the system in the world. 

We can now see how to ground the FEP in physical thermodynamics. In section (\ref{perceptron-thermo}), we saw that when the error probability is small, the change in the Helmholtz free energy per trial, for thermally activated perceptrons, is a minimum. This means that the work done on/by the agent at each new learning update is also small. This follows from the stochastic  thermodynamic fluctuation theorems~\cite{Seifert} that relate time averaged fluctuations in work done on/by the agent to changes in Helmholtz free energy of the agent. Once the agent has learned the right function to correctly predict its actions, it can control the world. The objective now switches to selecting actions (actuator settings) that result in particular sensations  (sensor records). This is determined by a \emph{policy}  function  $a=\pi(s)$. As this has already been learned by minimising the change in free energy per step, the correct policy will be one that minimises the change in free energy \emph{of the agent}. This is our version of the FEP.  

The EPL model just described bears some similarities to standard approaches to reinforcement learning (RL) in terms of a Markov decision process~\cite{Graesser}. 
The key difference between the agent-centered EPL model and the RL model is that the latter includes a description of states external to the agent, whereas the former is described only in terms of the internal sensor and actuator records of the agent, as well as internal variables (weights) governing the emulator. 

In RL as a Markov decision process, the agent's actions $a_t\in {\cal A}$ cause transitions in states $\sigma_t\in {\cal S}$ of the external world. There is no explicit reference to what the agent actually measures as it is simply assumed that the external states are perfectly perceptible, as is true of classical physics, but not true in the quantum case. For this reason we prefer a model in which actions cause transitions between internal states (sensors) of the learning agent itself.  In any case, the agent must learn an action-producing function $\pi\in{\cal P}$, called a policy, that maps states to actions $a=\pi(\sigma)$.

The full learning process now requires a specification of a model for the world in terms of how actions cause transitions between states. In a Markov model, this is given in terms of a transition probability $P(\sigma|\sigma',a')$ to determine how a  probability distribution of states of the world changes in response to actions and the prior state. We can think of $P(\sigma|\sigma',a')$ as the `law of physics' in the world.  At each iteration the world is supposed to provide a reward ${\cal R}$ that selects the policies $\pi$ that produce the most effective actions, for this reward, given  the `law of physics' in this world. In the ELM, the agent does not `know' this law but it can learn it by mimimizing the change in free energy per step of the learning feedback process.

In summary, EPL learns correlations inherent in actuator-sensor records in terms of the functions, $s=A(a)$, by minimising prediction errors and thus free energy, whereas RL learns policies that give actions as a function of states in the world, $a=\pi(\sigma)$ by maximising a reward. In EPL states of the world are `hidden', but if we assume that sensor records are analogous to the results of measurements made on the world, $s=M(\sigma)$, then  there is clearly a relation between the two models.

It is interesting to speculate at this point on why evolution should favour learning agents. An agent that spends too much time learning is going to consume a greater quantity of thermodynamic resources than one that has already learned a great deal about the class of functions $s=A_w(a)$ that it encounters. Furthermore, an agent that does not reach the optimum thermodynamic limit for learning will implement policies in control mode that waste thermodynamic resources.   Evolution will then favour learning agents.

\section{Discussion and Conclusion}
Learning machines are open, dissipative systems driven from thermal equilibrium. We have shown that the rate of the average energy dissipated, and average entropy, is proportional to the learning rate defined as the rate of change of the error probability. In all cases, learning requires an optimal level of noise. In the classical case, noise is due to thermal fluctuations. This implies that the learning rate decreases with temperature. In the quantum case, noise is due to quantum noise manifested in particular examples as spontaneous emission, quantum tunnelling or measurement back action noise.

The learning machines we have discussed are analogue devices in which dissipation and noise play a critical role. Can such systems exhibit  Turing universal computation? The subject of universal analogue computation has a long history going back at least to Shannon's work on general purpose analogue computation (GPAC)~\cite{GPAC}. A great deal of work has since been done to establish the universality of analogue computers~\cite{Pouly}. The approach is based on showing an equivalence between the solutions to differential equations and computational complexity classes of real functions. The conclusion is that analogue computation can be Turing universal. 

In the case of dissipative quantum learning machines we see another approach to harnessing quantum resources for computational purposes. This is in contrast to much of the work in quantum computation which is based on reversible unitary gate based models of computation. Recently however, a novel way of using quantum dissiptative systems to gain a computational advantage has emerged in coherent Ising machines~\cite{CIM}.  Clearly there is much more to be discovered in this area. 

Readers familiar with quantum computation may ask; what role does quantum entanglement play in a dissipative quantum learning machine?  The answer is that quantum tunnelling and spontaneous emission arise as a direct result of entanglement. In both cases, the rates depend on entanglement between local and global degrees of freedom following a quench: a local perturbation of the system. For example, a two level atom prepared in the ground state, and situated in a radiation vacuum state, is in a global stationary state. If the atom is subject to a local field --- a local quench --- that excites the  transition, it is no longer in a stationary state and rapidly evolves into an entangled state of the local electronic dynamics and the global radiation field. In the long time, this gives rise to spontaneous emission of a photon. In general, quantum noise in many-body systems arises due to entanglement between a local and global degree of freedom.   

When embedded in the environment, learning machines equipped with sensors and actuators are irreversible machines governed by the laws of thermodynamics. Embedded learning machines minimise error rate by minimising the rate of energy dissipation.  There is a considerable incentive to achieve this lower bound. In the case of biological systems, evolution should provide the incentive. In technology, economic constraints provide the incentive.   In both cases there is an incentive to make the most efficient use of thermodynamic resources.   Current implementations of learning via deep learning algorithms running on silicon based digital computers seem a long way short of this. 

There are a number of questions that arise in this perspective. Learning machines seek to capture the correlations between actions and sensations in a functional representation stored by the bias weights of the machine.  These functional relations are the `laws of physics' as far as the machine is concerned.  They can learn no more about the external world than are captured by the stored records in their sensors and actuators.  The physical implementation of sensors and actuators limit the laws that can be learned. This has implications for the metaphysics of physical law. One could argue that  technological prostheses enables science to  transcend the limitations of our biology. However, the hardware of our internal  learning machines necessarily limit what can even be imagined~\cite{Wolpert}.   

All the learning machines we are aware of are driven by thermal noise.  In biological systems, the learning machines are driven largely by thermally activated electro-chemical reactions.   At low temperature, quantum learning machines use quantum noise in place of thermal noise.  There would appear to be considerable scope here for innovations in learning machine technology.

\section*{Acknowledgments}
We would like to thank Natalia Ares, Andrew Briggs, Federico Fedele and Sally Shrapnel, for many valuable discussions.  This work is partly supported by Australian Research Council Centre of Excellence for Engineered Quantum Systems (EQUS, CE170100009). GJM acknowledges the support of Wolfson College, and the Department of Materials, Oxford University. S.B-E. acknowledges funding from the European Union Horizon 2020 research and innovation programme under the Marie Sk\l{}odowska-Curie grant agreement No 663830, also support from the S\^{e}r SAM Project at Swansea University, an initiative funded by the European Regional Development Fund through the Welsh Government’s S\^{e}r Cymru II Program.

\bibliography{learning}

\end{document}